\documentclass[preprint]{aastex}
\usepackage{amsmath} 

\newcommand{\myemail}{nathaniel.roth@berkeley.edu}
\newcommand{\Msun}{M_\odot}

\slugcomment{Accepted for publication in MNRAS}
\shorttitle{The Dynamics of Ultracompact HII Regions}
\shortauthors{Roth, Stahler, \&Keto}

\begin{document}

\title{The Dynamics of Ultracompact HII Regions}

\author{Nathaniel Roth\altaffilmark{1}}
\author{Steven W. Stahler\altaffilmark{2}}
\author{Eric Keto\altaffilmark{3}}
\email{\myemail}

\altaffiltext{1}{Dept. of Physics, U. of California, Berkeley, CA 94720}
\altaffiltext{2}{Dept. of Astronomy, U. of California, Berkeley, CA 94720}
\altaffiltext{3}{Harvard-Smithsonian Center for Astrophysics, 60 Garden Street, Cambridge, MA 02138}

\begin{abstract}
Many ultracompact HII regions exhibit a cometary morphology in radio continuum emission. In such regions, a young massive star is probably ablating, through its ultraviolet radiation, the molecular cloud clump that spawned it. On one side of the star, the radiation drives an ionization front that stalls in dense molecular gas. On the other side, ionized gas streams outward into the more rarefied environment. This wind is underpressured with respect to the neutral gas. The difference in pressure draws in more cloud material, feeding the wind until the densest molecular gas is dissipated.

Recent, time-dependent simulations of massive stars turning on within molecular gas show the system evolving in a direction similar to that just described. Here, we explore a semi-analytic model in which the wind is axisymmetric and has already achieved a steady state. Adoption of this simplified picture allows us to study the dependence of both the wind and its bounding ionization front on the stellar luminosity, the peak molecular density, and the displacement of the star from the center of the clump. For typical parameter values, the wind accelerates transonically to a speed of about 15~km~s$^{-1}$, and transports mass outward at a rate of $10^{-4}$ $M_\odot$ yr$^{-1}$. Stellar radiation pressure acts to steepen the density gradient of the wind.

\end{abstract}


\section{Introduction}
\subsection{Observational Background}

An ultracompact HII (UCHII) region is one of the earliest signposts for the presence of
a young, massive star (for reviews, see \citeauthor{Churchwell2002} 2002 and \citeauthor{Hoare2007} 2007). While the 
star itself is still too embedded in its parent molecular cloud to be detected
optically, it heats up surrounding dust. A small region, some $10^{17}$~cm in 
extent, glows brightly in the far infrared. Ionization of ambient gas also 
creates free-free emission in the radio continuum, with electron densities in excess of $10^4$ cm$^{-3}$ and thus emission measures of $10^7$ pc cm$^{-6}$ or more. It is through this radio emission, relatively minor in the overall energy budget, that UCHII regions have been classified morphologically. 

In their pioneering radio interferometric survey, \citet{Wood1989}
found that 30~percent of the spatially resolved regions have a cometary shape
(see also \citeauthor{Kurtz1994} 1994, \citeauthor{Walsh1998} 1998). One sees a bright arc of emission filled in by an extended, 
lower-intensity lobe that fades away from the arc. OH masers may be found along
the bright rim. Other UCHII regions exhibit only the emission arc; presumably the interior lobe in these cases is undetectably faint. Still other classes identified by \citet{Wood1989} include: spherical, core-halo (a bright peak surrounded by a fainter envelope), shell (a ring of emission), and irregular (multiple emission peaks). All told, the cometary morphology is the most common one found, and needs to be explained by any viable theoretical model. 

Spectral line studies have been used to probe the kinematics of these regions.
Observations of radio recombination lines \citep[e.g.][]{Afflerbach1996, Kim2001},
and infrared fine-structure lines \citep[e.g.][]{Zhu2005} reveal large 
line-widths, indicative of supersonic flow. In cases where the flow can be 
spatially resolved, one also sees a velocity gradient. This gradient is
steepest in the ``head-to-tail" direction \citep{Garay1994, Garay1999}. 

Very often, the observed peak in radio continuum or OH maser emission, either of which effectively locates the 
star, does {\it not} coincide with the peak in molecular lines or 
submillimeter continuum emission, which trace the densest molecular gas and dust \citep[e.g.][]{Mueller2002, Thompson2006}. This gas is located within infrared dark clouds, currently believed to be the birth sites of all massive stars \citep{Beuther2007}. The clumps within these clouds have typical sizes of 1 pc, number densities of $10^5$ cm$^{-3}$ and masses of about $10^4$ $M_\odot$; some qualify as hot cores \citep{Hofner2000, Hoare2007}. The offset of the peak radio emission of the UHCII region from the center of this clump is typically a few arcseconds, corresponding to approximately 0.1 pc for a distance of 1 kpc. Both the cometary morphology and the acceleration of ionized gas are likely related to this physical displacement, as was first emphasised by \citet{Kim2001}.

\subsection{Previous Models and Present Motivation}

The foregoing observations, taken together, show convincingly that the cometary
structures represent ionized gas accelerating away from the densest portion of 
the nearby cloud material. Theorists have long considered photoevaporating 
flows created by a massive star illuminating one face of a cloud 
\citep{Kahn1954, Oort1955}. The most well-studied classical HII region
of all, the Orion Nebula, is a prime example of the resulting ``blister,"
here formed by the massive star $\theta^1$~C on the surface of the Orion~A 
molecular cloud \citep{Zuckerman1973}. A tenuous, hemispherical body of ionized
gas surrounds $\theta^1$~C and the other Trapezium stars, and is flowing away
from the background cloud.

When massive stars are still deeply embedded in the densest portion of their parent cloud, it is not obvious how such photoevaporating flows can be maintained. \citet{Wood1989} pointed out that the small size of UCHII regions suggests a brief dynamical
lifetime. If they undergo pressure-driven expansion at $\sim 10$ km s$^{-1}$, they will expand to a size greater than 0.1~pc in roughly $10^4$~yr, or one percent of the lifetime of the host star. In reality, some 10 percent of O stars are associated with UCHII regions, suggesting that the lifetime of these regions is larger by an order of magnitude. Confinement by thermal pressure alone would result in an emission measure even higher than is observed \citep{Xie1996}. In the context of cometary structures, this venerable ``lifetime problem" raises a fundamental question. What reservoir of matter can feed the ionized flows over a period of $10^5$~yr?

\citet{Hollenbach1994} suggested that the star may be photoevaporating its own
accretion disk. When ultraviolet radiation from a massive star impinges on the disk, gas streams
off at the sound speed, at least in that outer region where this speed exceeds
the local escape velocity. \citet{Lugo2004} have analyzed this 
launching process in more detail.  The outer accretion disk radius of approximately $10^{15}$ cm is much smaller than the $10^{17}$~cm size of UCHII regions. Thus, while the model views a disk as the ultimate source of matter for the ionized flow, it does not address the flow's cometary morphology. 

One possibility is that the prominent arc 
represents the shock interface between a high-velocity stellar wind and the 
parent cloud. Massive stars indeed generate, through radiative acceleration, 
winds with terminal velocities of about 1000~km~s$^{-1}$. If the star itself 
moves through the cloud at supersonic speed, e.g. 20~km~s$^{-1}$, then the 
curved bowshock has the right form \citep{Vanburen1990, Mac-Low1991}. Moreover, a star that is moving toward the 
observer creates a ``core-halo" structure, also commonly seen. The relatively 
fast stellar velocity, however, implies that the whole interaction would last for a few times $10^4$ years if the star indeed travels through a molecular clump 1 pc in size. Consequently, this bowshock model may fall prey to the lifetime problem. Another concern is that massive stars, with the exception of runaway objects, do not move at such high speeds relative to parent molecular gas. For instance, the Trapezium stars in the Orion Nebula Cluster have a velocity dispersion of a few km s$^{-1}$ \citep{Furesz2008}.

Suppose instead that the embedded star is not moving with respect to the densest gas, but is displaced from it, as is suggested by the observations. Then one side of its expanding HII region eventually erupts into the surrounding low-density medium. Such a ``champagne flow'' model was first explored by \citet{Tenorio-Tagle1979}, \citet{Bodenheimer1979}, and \citet{Whitworth1979} to explain asymmetric, classical HII regions. The high internal pressure of the HII region accelerates the ionized gas to supersonic speed away from the ambient cloud, whose density was taken to be $10^3$ cm$^{-3}$. Meanwhile, the ionization front steadily advances at several km~s$^{-1}$ into this cloud, generating volumes of ionized gas several parsecs in diameter.

Another model, that of the mass-loaded wind \citep{Dyson1968}, invokes champagne-type dynamics in combination with a stellar wind. The idea is that the wind, perhaps in combination with the stellar radiation field, ablates the cloud, entraining the gas within it. Pressure gradients may then accelerate the gas in a champagne flow. In some versions of the model, the new mass originates in dense globules that are continually being ionized \citep{Williams1996,Lizano1996,Redman1998}. While successful in explaining the morphologies and lifetimes of UCHII regions, the model does not address in detail how the dense globules enter the ionized flow. Thus, the mass-loading prescription adopted was somewhat ad hoc.

\citet{Keto2002-2} modeled the growth of an HII region inside a Bondi accretion flow. If the ionization front is located inside the Bondi radius  $r_B \equiv GM_\ast/2 a_I^2$, where $M_\ast$ is the stellar mass and $a_I$ the ionized sound speed, then gas simply crosses the ionization front and continues toward the star as an ionized accretion flow. Until its size reaches $r_B$, the HII region can expand only as the ionizing flux from the star increases along with the stellar mass. The HII region then exists in steady state, continuously fed by the molecular accretion. Thus, the lifetime of the HII region is tied to the accretion time scale of the star. 

Observations of the UCHII region around the cluster of massive stars G10.6-0.4  indicate that this kind of molecular and ionized accretion could be occurring \citep{Keto2002-1}. Subsequent observations of the same cluster \citep{Keto2006} also show an asymmetric bipolar outflow of ionized gas. These observations motivated \citet{Keto2007} to develop a model in which inflow and outflow occur simultaneously in a rotationally flattened geometry. The ionization creates an HII region elongated perpendicular to the accretion flow. Where the ionization extends beyond $r_B$, the HII region can expand hydrodynamically as a pressure-driven Parker wind (see also \citet{McKee2008}). Along the equatorial plane, dense molecular gas continues to flow into the HII region.  One of the motivations of the present paper is to detail exactly how an ionized outflow may be supplied by inflow of molecular gas, although here we consider a mechanism to draw in this gas that is separate from the gravitational attraction of the star.

Another model combining ionization and gravitational accretion was that of \citet{Mac-Low2007}, who suggested that an UCHII region may form as a gravitationally collapsing substructure within a larger, expanding HII region. More detailed simulations by \citet{Peters2010-1, Peters2010-2} found that an UCHII region embedded in an accretion flow rapidly changes morphologies through all the observed types, and could be sustained by the addition of infalling gas from the parent cloud. The collapsing ionized gas in their simulations creates bipolar molecular outflows, as are observed to accompany some UCHII regions \citep{Beuther2005}. \citet{Peters2011} also claimed that magnetic pressure might play a significant role in confining UCHII regions. On the other hand, \citet{Arthur2011} performed their own radiation-MHD simulation of an HII region expanding into a turbulent cloud. They found that magnetic pressure plays only a minor role in confinement.

Other recent simulations have combined various elements in the original models to account for the proliferating observational results now becoming available. \citet{Henney2005} simulated a champagne flow in which the ionization front is stalled as it climbs a density gradient into a neutral cloud supported by turbulent pressure. \citet{Comeron1997} explored the coupled roles of champagne flows and stellar winds. Finally, \citet{Arthur2006} revised the bowshock idea by introducing a finite but modest stellar velocity with respect to the cloud.
 
In the present paper we pursue a different goal. Accepting that every cometary UCHII region is an ionized flow, we elucidate the basic issue of how such a flow may continue to draw in gas from the neutral cloud. We agree with previous researchers that the momentum of the ionized flow is not supplied by the star, but by the thermal pressure gradient of the gas itself. A key result of our own study will be to demonstrate that the expansion of this flow causes the ionized gas to become {\it underpressured} with respect to the neutral cloud. This pressure differential draws neutral material from the cloud into the champagne flow in a self-sustaining manner.

We assume at the outset that the flow is quasi-steady, as was found in the late stages of the simulations of \citet{Henney2005} and \citet{Arthur2006}. This basic simplification allows us to rapidly explore parameter space, as we detail below. Thus, we can assess how well our rather minimal set of physical assumptions explains the basic characteristics of cometary UCHII regions.  Our quasi-one dimensional model does not allow us to include evacuation by a stellar wind, but we do account for the effect of radiation pressure and show that it is appreciable. 

In Section~\ref{ModelDescription} below, we introduce our steady-state model for cometary UCHII regions. Section~\ref{FlowEquations} develops the equations governing the density and velocity of the ionized flow. In Section~\ref{Nondimensionalization}, we recast these equations in nondimensional form and then outline our solution strategy. Section~\ref{Results} presents our numerical results, and Section~\ref{ComparisonToObservations} compares them to observations. Finally, Section~\ref{Conclusion} indicates fruitful directions for future investigations.

\section{Steady-state Model}
\label{ModelDescription}
\subsection{Physical Picture}

\begin{figure}
\plotone{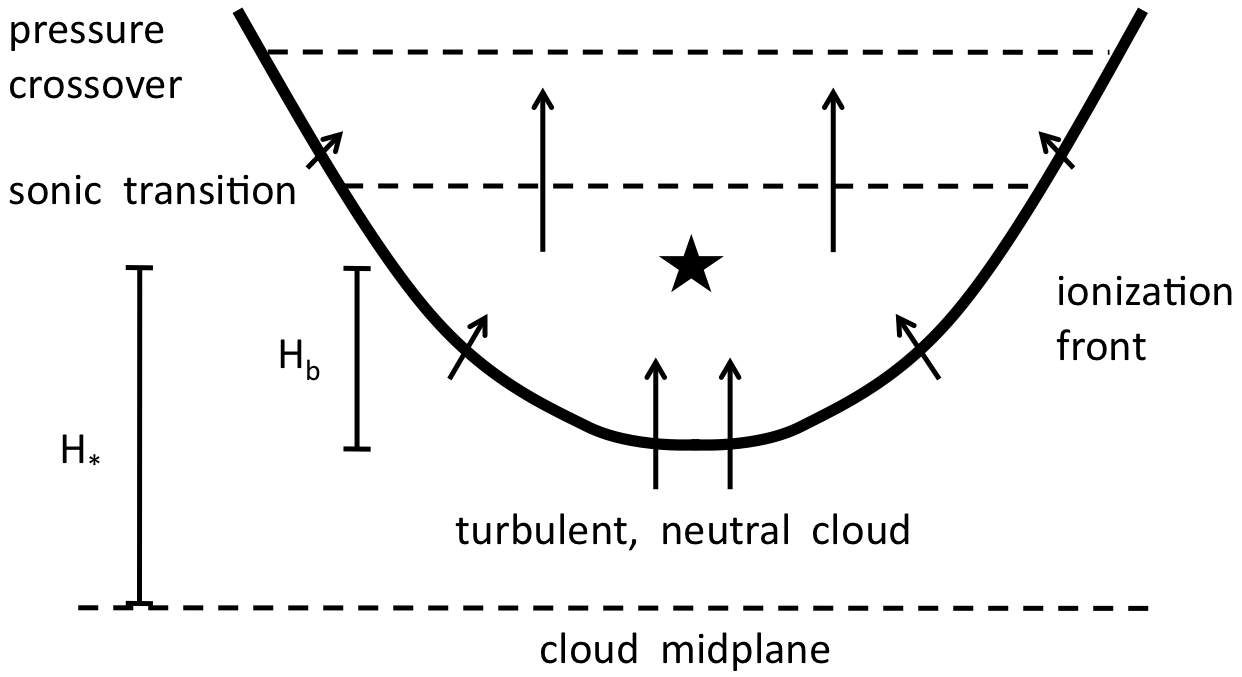}
\caption{Our flow schematic. A massive star is situated to one side of an initially neutral cloud. It creates, as in the original champagne model, an asymmetric HII region of relatively low density. 
Toward the midplane, the ionization front is stalled by rising cloud density. 
In the opposite direction, the front has broken free. The high pressure of 
ionized gas creates an accelerating flow away from the densest gas. At the base of the ionized flow, the pressure has been relieved sufficiently such that the neutral cloud gas is slightly overpressured with respect to the ionized gas.}
\label{schematic}
\end{figure}

We idealize the molecular cloud as a planar slab, in which the cloud density peaks at the midplane (see Figure \ref{schematic}). 
The choice of planar geometry is made for computational convenience and is probably not a realistic representation of the clouds of interest. However, our qualitative results are not sensitive to the adopted cloud density profile, as we verify explicitly in Section \ref{Results} below. We envision a massive star embedded within this cloud, but offset from the midplane, in accordance with the observations mentioned previously. This offset creates an asymmetric, ionized region of relatively low density. Toward the cloud midplane, this ionized gas resembles a classical HII region. Its boundary, a D-type ionization front, advances up the density gradient. In the opposite direction, the front breaks free and gas streams away supersonically. The pressure gradient within the 
ionized gas creates an accelerating flow away from the cloud. Apart from the higher-density cloud environment, our model is identical in spirit to the champagne flows proposed in the past.

Once the velocity of the advancing ionization front falls significantly below the ionized sound speed, the flow becomes steady-state. We adopt this steady-state assumption in our model, and assume that the structure of the background cloud is evolving over a long timescale compared to the ionized sound crossing time. In the simulations of \citet{Arthur2006}, a steady flow is approached some $10^5$ years after the massive star turns on. By this point, the ionization front has virtually come to rest in the frame of the star. We will later determine more precisely the speed of the ionization front in our model and show that it is consistent with the steady-state assumption. The shock that originally formed ahead of the ionization front, as it transitioned from R-type to D-type, has by now advanced deep into the cloud and died away. 

The HII region facing the cloud midplane expands, albeit slowly. While the ionized gas is optically thick to ultraviolet radiation from the star, some does leak through and strikes the cavity wall within the neutral cloud. Additional gas is thus dissociated and ionized, and streams off the wall to join the outward flow. While the injection speed at the wall is subsonic, the thermal pressure gradient within the ionized gas accelerates it to supersonic velocity. Both this flow and the advancing front erode the cloud, whose structure gradually evolves in response. 

\subsection{Tracing the Ionization Front}

We will always consider systems that possess azimuthal symmetry. We establish a spherical coordinate system whose origin is at the star. The polar direction, $\theta = 0$, coincides with the central axis of the flow depicted in Figure \ref{schematic}. Let ${\cal N}_\ast$ denote the total number of ionizing photons per time generated by the star. If we further assume
ionization balance within the volume of the ionized cavity, and make use of the on-the-spot approximation, then the ionizing radiation extends out to the cavity wall
$r_f(\theta)$, given implicitly by 
\begin{equation}
{{{\cal N}_\ast}\over{4\,\pi}} \,=\, 
\alpha_B\,\int_0^{r_f}\!dr\,r^2\,[n_I(r,\theta)/2]^2 \,+\, r_f^2\,F_{\ast \, , \, {\rm wall}} (\theta) \,\,.
\label{ExactRayTracing}
\end{equation} 
Here, $\alpha_B$ is the case B recombination coefficient (Osterbrock \citeyear{Osterbrock1989} Chapter 2). We have pulled
this factor out of the integral since it depends only on the ionized gas temperature, 
which we assume to be spatially uniform. The term $n_I$ is the number density of ionized gas, including both protons and electrons. For simplicity, we have assumed that the composition of the gas is pure hydrogen, so the number of free protons is equal to the number of free electrons and both of these number densities are equal to $n_I/2$. Finally, $F_{\ast \, , \, {\rm wall}}(\theta)$ is that portion of the star's ionizing photon flux which escapes the HII region and strikes the wall of the cavity. This remnant flux is critical for maintaining the flow. 

But how important is the second term of equation (\ref{ExactRayTracing}) in a quantitative sense? Each photon striking the front ionizes a hydrogen atom, itself already dissociated from a hydrogen molecule, thereby creating a proton and an electron. Let $f$ be the number flux of protons and electrons injected into the flow at each position along the front. Thus, $F_{\ast \, , \, {\rm wall}} = f/2$ at each angle $\theta$, and we may write the foregoing equation as
\begin{equation}
{{{\cal N}_\ast}\over{4\,\pi}} \,=\,
\alpha_B\,\int_0^{r_f}\!dr\,r^2\,(n_I/2)^2 \,+\, {{r_f^2\,f}\over 2} \,\, .
\label{RayTracingRewritten}
\end{equation}

The flux $f$ is of order $n_I\,a_I$, where $n_I$ is now a typical value of the
ionized gas number density. If $r_f$ in equation (\ref{RayTracingRewritten}) now represents the size scale of the flow, then the ratio of the second to the first righthand term in this equation is of order
\begin{equation}
{{a_I}\over{\alpha_B\,n_I\,r_f}} \,\sim\, 
\left({{a_I}\over{a_{\rm cl}}}\right)^3
{a_{\rm cl}\over{\alpha_B\,n_{\rm cl}\,r_f}}
 \,\, ,
\end{equation}
where $a_\mathrm{cl}$ and $n_\mathrm{cl}$ are the effective sound speed and number density within the cloud (see below). Here we have assumed that the ionized material is at least in rough pressure balance with the cloud, so that \hbox{$n_I\,a_I^2 \approx n_{\rm cl} \,a_{\rm cl}^2$}. For the dense clumps that harbor young massive stars, 
\hbox{$n_{\rm cl}\,\sim\,10^5\ {\rm cm}^{-3}$} and
\hbox{$a_{\rm cl}\,\sim\,2\ {\rm km}\,{\rm s}^{-1}$} \citep{Garay1999}. Further 
using \hbox{$r_f\,\sim\,10^{17}\ {\rm cm}$},
\hbox{$a_I\,\sim\,10\ {\rm km}\ {\rm s}^{-1}$}, and
\hbox{$\alpha_B\,\sim\,10^{-13}\,\,{\rm cm}^3\,{\rm s}^{-1}$}, we find that the
ratio of terms is of order $10^{-2}$. In practice, therefore, we neglect the
second term entirely.\footnote{In the terminology of \citet{Henney2001}, our 
photoevaporation flow is {\it recombination-dominated}. If the second 
righthand term in equation~(\ref{RayTracingRewritten}) were relatively large, the flow would be 
{\it advection-dominated}. According to \citet{Henney2001}, knots in planetary 
nebulae fall into this category.} We trace the ionization front by finding that 
function $r_f(\theta)$ which obeys the approximate, but quantitatively accurate relation,
\begin{equation}
{{{\cal N}_\ast}\over{4\,\pi}} \,=\, \alpha_B\,\int_0^{r_f}\!dr\,r^2\,(n_I/2)^2 \,\,.
\label{RayTracingApproximate}
\end{equation}

\subsection{Cloud density and gravitational potential}
\label{DensityAndPotential}

Even within our simplifying assumption of axisymmetry, it would be a daunting task to trace a fully two-dimensional flow. While we accurately follow the ionization front bounding the flow in two dimensions, we further assume that the density and velocity are only functions of $z$. In this quasi one-dimensional model, we are effectively averaging the ionized density $n_I$ and velocity $u$ laterally at each $z$-value. This simplification is innocuous in regions where the lateral change in these quantities is relatively small. It is more problematic when the ionized gas becomes overpressured with respect to the ambient cloud. Luckily, most of the radio emission comes from the densest portion of the flow, well before this point is reached. Hence, our model produces reasonably accurate results when predicting observed emission measures. 

Turning to the cloud itself, we assume it to be self-gravitating, with an internal supporting pressure arising from turbulence. Hence, we do not consider the possibility that the object is in a state of collapse. We crudely model the turbulence by adopting an isothermal equation of state, characterized by an effective sound speed $a_{\rm cl}$. In our model, it is only the mass of the cloud that affects the flow gravitationally. That is, we neglect both the self-gravity of the ionized gas, and the pull of the massive star. The latter force is negligible outside the Bondi radius \hbox{$R_B \,\equiv\,G\,M_\ast/2\,a_I^2$}, where $M_\ast$ is the
stellar mass. For representative values \hbox{$M_\ast\,=\,20\ \Msun$} and
\hbox{$a_I\,=\,10\ {\rm km}\,{\rm s}^{-1}$}, \hbox{$R_B \,=\, 90\ {\rm AU}$},
much less than the size of UCHII regions.

Our cloud is an isothermal self-gravitating slab whose midplane is located at $z = -H_{\ast}$ (see Fig. 1). Solving the equations of hydrostatic equilibrium along with Poisson's equation yields the neutral cloud density $n_{\rm cl}$ and the gravitational potential $\Phi_{\rm cl}$
\begin{equation}
n_{\rm cl} \,=\, n_0\,{\rm sech}^2\,
\left({{z\,+\,H_\ast}\over H_{\rm cl}}\right) \,\,,
\label{SlabDensity}
\end{equation}
\begin{equation}
\Phi_{\rm cl} = -a_{\rm cl}^2\,\,{\rm ln}\,{\rm sech}^2
\left({{z\,+\,H_\ast}\over H_{\rm cl}}\right) \,\,.
\label{SlabPotential}
\end{equation}
Here, $n_0$ is the midplane number density of hydrogen molecules, each of mass $2 \, m_H$, and $H_{\rm cl}$ is the scale thickness of the slab:
\begin{equation}
H_{\rm cl} \,\equiv\, 
\left[{a_{\rm cl}^2\over{2\,\pi\,G\,(2 \, n_0 \, m_\mathrm{H})}}\right]^{1/2} \,\,.
\label{ScaleHeight}
\end{equation}

A representative value for $n_0$ is $10^5$ cm$^{-3}$. Combining this with a turbulent velocity dispersion of $a_{\rm cl} = 2 $ km s$^{-1}$ yields a Jeans mass on the order of $10^2 \, M_\odot$, far less than the $10^{4} \, M_\odot$ clumps in giant molecular clouds spawning massive stars and their surrounding clusters. We stress again that our cloud is a relatively small fragment containing the newborn star. In our view, it is the high density of this gas that determines the morphology of the UCHII region, and the dispersal of the clump that sets the characteristic lifetime.

\subsection{Radiation pressure}
\label{RadPressureExplained}
Quantitative modeling of HII regions, and UCHII regions in particular, has generally neglected the dynamical effect of radiation pressure from the massive star (see, however, \citeauthor{Krumholz2009} \citeyear{Krumholz2009}). For the very dense environments we are now considering, the radiative force has substantial influence on the ionized flow, as we shall demonstrate through explicit calculation.

Suppose the star emits photons with mean energy $\epsilon.$ Those traveling in the direction $\theta$ with respect to the central axis carry momentum $(\epsilon / c ) \cos \theta$ in the $z$-direction. If we assume that the gas is in ionization equilibrium, then the number of photons absorbed per unit volume of gas equals the corresponding volumetric rate of recombination, $(n_I/2)^2 \alpha_B$. The assumption of ionization equilibrium is justified because the typical recombination time, $t_{\rm rec} \sim \left({n_I \alpha_B}\right)^{-1} \sim $ 10$^8$ s, is much less than the flow time $t_{\rm flow} \sim r_f/a_I \sim $ 10$^{11}$ s. 

The radiative force per volume is the product of the photon momentum and the volumetric rate of ionization, or $\left( \epsilon / c \right)\cos{\theta} \, (n_I/2)^2 \alpha_B $. To obtain $f_{\rm rad}$, the radiative force per unit \emph{mass} of gas, we divide this expression by the ionized mass density, $m_H n_I / 2$. We thus find 

\begin{equation} 
f_\mathrm{rad} = \frac{\epsilon \,\langle \cos \theta \rangle \, n_I \, \alpha_B}{2 \, m_H c} \, \, .
\label{FradDefinition}
\end{equation}
This expression is consistent with that given in \citet[][Section 2]{Krumholz2009} if we add the condition of ionization balance. Note, finally, that $\epsilon$ in equation (\ref{FradDefinition}) is implicitly a function of ${\cal N}_{\ast}$, a fact that we shall use later.

In accordance with our quasi one-dimensional treatment, we have laterally averaged $\cos \theta$ at fixed $z$. Explicitly, this average, weighted by the cross-sectional area is 

\begin{equation}
\langle \cos \theta \rangle = 2 \, \eta \, \frac{z^2}{R^2} \left(\sqrt{1 + \frac{R^2}{z^2}} - 1 \right) \, \, .
\label{AverageCosine}
\end{equation}
Here, $R(z)$ is the cylindrical radius from the central axis to the ionization front at each $z$, and $\eta = +1$ or $-1$ for positive and negative $z$, respectively. At \hbox{$z = 0$}, the level of the star, we set \hbox{$\langle \cos \theta \rangle= 0$}. At the base of the flow, \hbox{$\langle \cos \theta \rangle= -1$}. We define the distance from this point to the star as $H_b \equiv r_f(\pi)$, and show it in Figure \ref{schematic}. 

\section{Flow equations}
\label{FlowEquations}

\subsection{Mass and Momentum Conservation}

\begin{figure}
\plotone{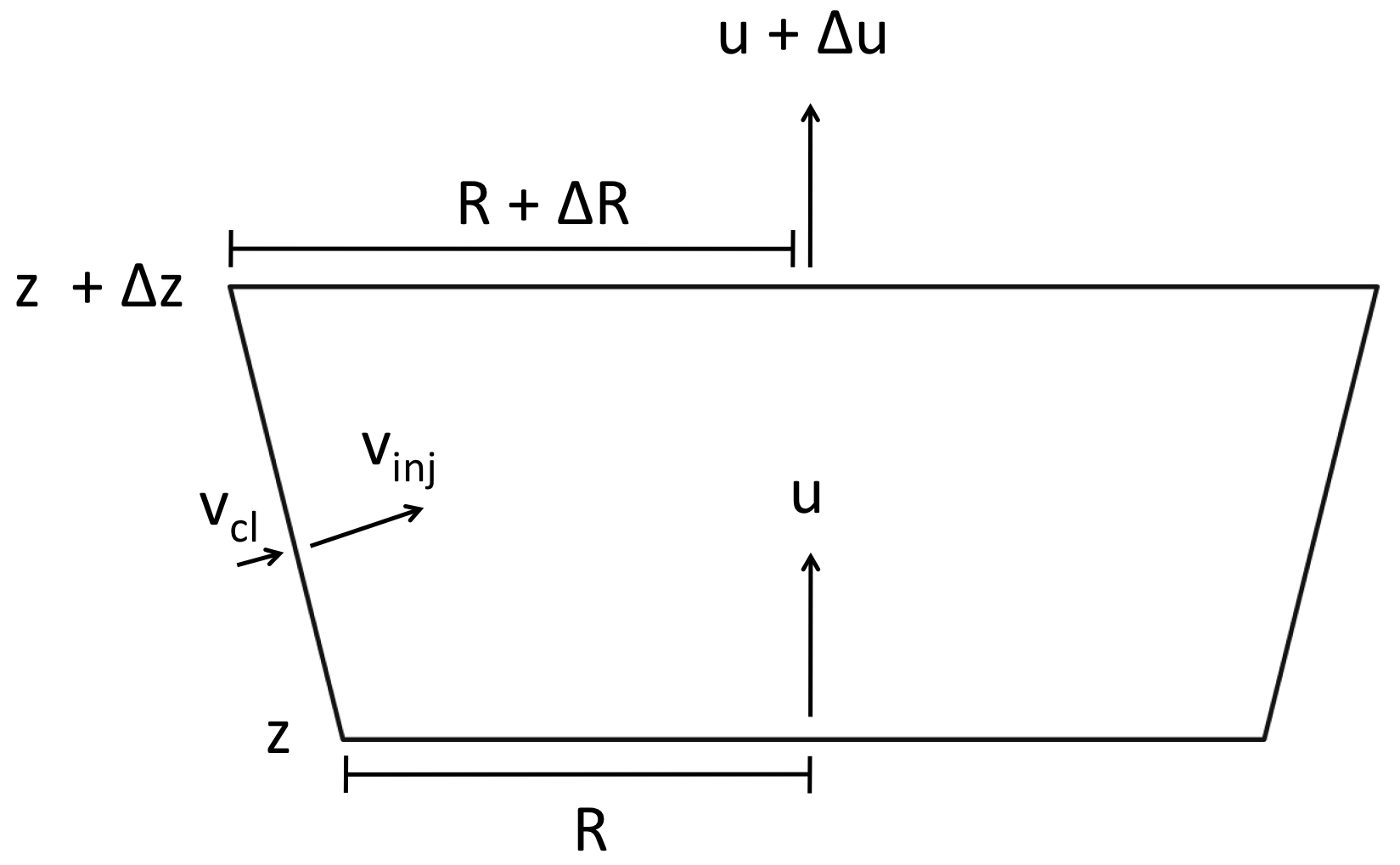}
\caption{Control volume diagram for a segment of the ionized flow. The trapezoidal section displayed here represents a meridional slice of the ionised flow - one should imagine rotating this diagram about the central vertical symmetry axis in order to generate the represented volume.}
\label{ControlVolume}
\end{figure}

To obtain laterally averaged dynamical equations, consider a control volume of height $\Delta z$ spanning the flow, as pictured in Figure \ref{ControlVolume}. This volume has cylindrical radius $R$ and $R + \Delta R$ at its lower and upper surfaces, respectively. Let $u_I(z)$ be the average flow speed in the $z$-direction. Similarly, let $n_{\rm inj}$ and $v_{\rm inj}$ represent the number density and speed, respectively, of ionized gas being injected into the flow just inside the ionization front. Then the requirement of mass conservation is
\begin{equation}
\frac{d}{dt}(\pi\, R^2 \, \Delta z \, n_I) = \pi \, R^2 \, n_I \, u_I - \pi (R+\Delta R)^2 (n_I + \Delta n_I) (u_I + \Delta u_I) + 2 \, \pi \, R \,\sqrt{\Delta R^2 + \Delta z^2} \, n_{\rm inj} \, v_\mathrm{inj} \, \, . 
\label{FirstMassConservation}
\end{equation}
Here, the first two terms of the righthand side are the rate of mass advection through the bottom and top layers of the control volume, respectively. The final righthand term is the rate of mass injection. We assume that the injected flow direction is normal to the ionization front. Henceforth, we will drop the subscript $I$ when referring to the density and velocity of the ionized flow.

Under our steady-state assumption, the lefthand side of equation (\ref{FirstMassConservation}) vanishes. Dividing through by $\Delta z$, and taking  the limit $\Delta z \rightarrow 0$, leads to 
\begin{equation}
\frac{d}{dz}(R^2 \, n \, u) = 2 \, R \, \sqrt{1 + \left(\frac{dR}{dz}\right)^2}  n_{\rm inj} \, v_\mathrm{inj} \, \, . 
\end{equation}
We will later derive an expression for $v_\mathrm{inj}$ from the jump conditions across the ionization front and show that this velocity is subsonic, i.e., $v_\mathrm{inj} < a_I$ at any $z$. Although our derived $v_\mathrm{inj}$ is properly measured in the rest frame of the ionization front, we will show in Section \ref{JumpConditions} that the front is moving very slowly compared to $a_I$. Thus, $v_\mathrm{inj}$ is also, to a high degree of accuracy, the speed of the injected gas in the rest frame of the star.

Another crucial assumption we will make is that $n_\mathrm{inj} = n$, i.e., the injected density is the same as the laterally averaged flow value at any height. This condition seems to hold at least approximately in the simulations of \citet{Arthur2006}, and is physically plausible when one considers that lateral density gradients will tend to be smoothed out if the velocity components in that direction are subsonic. After making this assumption, we are left with
\begin{equation}
\frac{d}{dz}(R^2 \, n \, u) = 2 \, R \, \sqrt{1 + \left(\frac{dR}{dz}\right)^2}  n \, v_\mathrm{inj} \, \, .
\label{MassConservation}
\end{equation}

Requiring conservation of $z$-momentum for the control volume leads to 
\begin{align}
\frac{d}{dt}(\pi\, R^2 \, \Delta z \, n \, u) &= \pi \, R^2 \, n \, u^2 - \pi (R+\Delta R)^2 (n+ \Delta n) (u + \Delta u)^2 + 2 \, \pi \, R \, n \, v_\mathrm{inj}^2 \,\Delta R \nonumber \\
&+ \pi \, R^2 \, n \,{a_I}^2 - \pi \, (R + \Delta R)^2 (n + \Delta n) {a_I}^2 + 2 \, \pi \, R \, \Delta R^2 + \Delta z^2 \, n \, {a_I}^2 \Delta R \nonumber \\ &- \pi \, R^2 \, \Delta z \, n \frac{d\Phi_{\rm cl}}{dz} + \pi \, R^2 \, \Delta z \, n \, f_\mathrm{rad} \,\, ,
\label{ControlVolumeMomentum}
\end{align}
where $\Phi_{\rm cl}$ is the gravitational potential from equation (\ref{SlabPotential}) and $f_\mathrm{rad}$ is the radiative force per mass from equation (\ref{FradDefinition}). Included are terms representing both static pressure and the advection of momentum through the top, bottom, and sides of the control volume. For the advective terms, we have again replaced $n_{\rm inj}$ by $n$. The geometric factor $\sqrt{1 + (d R/d z)^2}$ present in equation (\ref{MassConservation}) disappears because its inverse is used when projecting the injected momentum into the $z$-direction. We apply the steady-state condition and divide equation (\ref{ControlVolumeMomentum}) through by $\pi R^2 \, n \, \Delta z$. After taking the $\Delta z \rightarrow 0$ limit and combining with equation (\ref{MassConservation}), we obtain 
\begin{equation}
u \, \frac{du}{dz} = - \frac{a_I^2}{n} \, \frac{dn}{dz}  - \frac{d\Phi_{\rm cl}}{dz} + f_\mathrm{rad} + \frac{2}{R} \, \frac{dR}{dz} \, v_\mathrm{inj}^2  - \frac{2}{ R} \, \sqrt{1 + \left(\frac{dR}{dz}\right)^2} \, v_\mathrm{inj} \, u \, .
\label{MomentumConservation}
\end{equation}
This equation resembles the standard Euler momentum equation in one dimension, but has two additional terms. The first accounts for injection of $z$-momentum via ram pressure. The second represents the inertial effect of mass loading.

\subsection{Jump conditions across the ionization front}
\label{JumpConditions}
In the rest frame of the ionization front, upstream molecular gas approaches at speed $v_\mathrm{cl}^\prime$ and leaves downstream as ionized gas, at speed $v_\mathrm{inj}^\prime$. We assume that the intermediate photodissociation region, consisting of neutral hydrogen atoms, is geometrically thin.\footnote{\citet[Section 4.2]{Roshi2005} estimate the PDR thickness in G35.20-1.74 to be of order $10^{-4}$ pc.} Conservation of mass and momentum across the ionization front is expressed in the jump conditions 
\begin{mathletters}
\begin{eqnarray}
\label{JumpCondition1}
(1/2) \, n \, v_\mathrm{inj}^\prime &=& 2 \, n_\mathrm{cl} \, v_\mathrm{cl} ^\prime  \\
(1/2) \, n \, (a_I ^2 + {v_\mathrm{inj}^\prime}^2) &=& 2 \, n_\mathrm{cl}\, (a_\mathrm{cl}^2 + {v_\mathrm{cl}^\prime}^2) \, .
\label{JumpCondition2}
\end{eqnarray}
\end{mathletters}
The factors of $1/2$ and 2 in both equations account for the fact that each hydrogen molecule has a mass of $2 \, m_H$, while each particle of ionized gas has a mean mass of $m_H/2$. We solve equation (\ref{JumpCondition1}) for ${v_\mathrm{cl}^\prime}$
\begin{equation}
v_\mathrm{cl}^\prime  = \frac{1}{4} \frac{n}{n_\mathrm{cl}} v_\mathrm{inj}^\prime \, \, ,
\label{CloudVelocity}
\end{equation}
and use this result in equation (\ref{JumpCondition2}) to derive an expression for $v_\mathrm{inj}^\prime$:

\begin{equation}
v_\mathrm{inj}^\prime = a_I \sqrt{\frac{16 n_\mathrm{cl}^2/\beta - 4 n \, n_\mathrm{cl}}{4 n\,n_\mathrm{cl} - n^2}} \, \, ,
\label{VinjPrime}
\end{equation}
where \hbox{$\beta \equiv a_I^2/a_\mathrm{cl}^2 \gg 1$}. As long as the ionized gas is underpressured with respect to the neutral molecular gas, \hbox{$\beta \, n < 4  \, n_{\rm cl}$}, and the quantity inside the square root in equation (\ref{VinjPrime}) is positive, guaranteeing a solution for $v_{\rm inj}^\prime$. Using this inequality, equation (\ref{CloudVelocity}) tells us that \hbox{$v_{\rm cl}^\prime < v_{\rm inj}^\prime/ \beta$ }. In all our solutions, $v_{\rm inj}^\prime$ is subsonic. Thus, the ionization front moves relatively slowly into the cloud, and we may set the lab frame injection velocity $v_\mathrm{inj}$ equal to $v_\mathrm{inj}^\prime $. Since we know $n_\mathrm{cl}$ at all $z$, equation (\ref{VinjPrime}) gives us $v_{\rm inj}$ as a function of $z$ and $n$, which we may use in the mass and momentum equations (\ref{MassConservation}) and (\ref{MomentumConservation}).

\subsection{Decoupled equations of motion}

The mass and momentum conservation equations can be combined to solve separately for the derivatives of the velocity and density. These decoupled equations are
\begin{equation}
\label{DecoupledEquations1}
\frac{du}{dz} = \left(\frac{1}{a_I^2 - u^2}\right)\left[-2 \,u \, \frac{1}{R} \, \frac{dR}{dz} (a_I^2  + v_\mathrm{inj}^2) + 2 \, v_\mathrm{inj} \frac{1}{R}\,\sqrt{1 + \left(\frac{dR}{dz}\right)^2}\, (a_I^2 + u^2)  + u \frac{d\Phi_{\rm cl}}{dz}  - u \, f_\mathrm{rad} \right]
\end{equation}
\begin{equation}
\label{DecoupledEquations2}
\frac{dn}{dz} = \left(\frac{1}{a_I^2 - u^2}\right)\left[2 \, n \, \frac{1}{R} \, \frac{dR}{dz} (u^2 + v_\mathrm{inj}^2)  - 4 \, n \, \frac{1}{R}\, \sqrt{1 + \left(\frac{dR}{dz}\right)^2} \,u \,v_\mathrm{inj} - n \frac{d\Phi_{\rm cl}}{dz}  + n \, f_\mathrm{rad} \right] \, \, .
\end{equation}

The righthand sides of both equations have denominators that vanish when $u = a_I$. Thus, we must take special care when integrating through the sonic point, as is also true in steady-state winds and accretion flows of an isothermal gas. If we were to ignore the terms relating to mass injection, radiation pressure, and self-gravity, then the sonic transition would occur when $dR/dz = 0$, as in a de Laval nozzle. In our case, the extra terms cause the sonic transition to occur at other locations.

\section{Nondimensionalization and solution strategy}
\label{Nondimensionalization}
\subsection{Characteristic scales}

A represenative ionizing photon emission rate is $10^{49}$ s$^{-1}$, which corresponds to an O7.5 star \citep{Vacca1996}. We denote this emission rate by ${\cal N}_{49}$. We define a nondimensional emission rate normalized to that value:
\begin{equation}
\tilde{{\cal N}}_\ast = \frac{{\cal N}_\ast}{{\cal N}_{49}} \, \, .
\end{equation}

To find characteristic density and length scales for the flow, consider first $H_{\rm{cl}}$, the scale height of the neutral cloud. According to equation (\ref{ScaleHeight}), this quantity depends on both the effective sound speed $a_\mathrm{cl}$, which we fix at 2 km/s, and on the midplane cloud density $n_0$, which will be a free parameter. A second length of importance is the Str\"omgren radius of fully ionized gas of uniform particle number density $n$, given by   
\begin{equation}
R_S \equiv \left(\frac{3 {\cal N}_\ast}{4 \pi \alpha_B (n/2)^2} \right)^{1/3} \, \, .
\label{StromgrenRadius}
\end{equation}
Notice the appearance of $n/2$ in our expression. This is the number density of either protons or electrons, the species that actually recombine.

For a cometary flow to exist at all, it must be true that $R_S \sim H_\mathrm{cl}$. If $R_S \ll H_\mathrm{cl}$, the HII region would be trapped within the cloud and unable to generate the observed flow. If, on the other hand, $R_S \gg H_\mathrm{cl}$, the ionized region would be free to expand in all directions, again contrary to observation. For the purpose of defining a characteristic ionized density scale, we first set ${\cal N}_\ast = {\cal N}_{49}$ in equation (\ref{StromgrenRadius}). We then set $n_0 = \beta n /4$ in equation (\ref{SlabDensity}), and solve for the ionized density $n$ that satisfies the relation $H_\mathrm{cl} = R_S$. We label this density $n_{49}$, and find that it can be expressed as 
\begin{equation}
n_{49} = 9 \pi \,\beta^3 \left(\frac{G \, m_H}{a_\mathrm{cl}^2}\right)^3 \left( \frac{{\cal N}_{49}}{\alpha_B}\right)^2 = 1.4 \times 10^4 \; \mathrm{cm}^{-3} \, \, .
\label{FiducialDensity}
\end{equation}
We insert this value of $n_{49}$ into equation (\ref{StromgrenRadius}) and set the resulting $R_S$ equal to the characteristic length scale $Z_{49}$. We find for this length

\begin{equation}
Z_\mathrm{49} = \frac{1}{3 \pi \, \beta^2} \left(\frac{G \, m_H}{a_\mathrm{cl}^2}\right)^{-2} \left( \frac{{\cal N}_{49}}{\alpha_B}\right)^{-1} = 0.18 \; \mathrm{pc} \, \, .
\label{FiducialLength}
\end{equation}
It is encouraging that our values for $n_{49}$ and $Z_{49}$ match typical observations for UCHII regions \citep{Churchwell2002}. 

\subsection{Nondimensional equations}

We first normalize all lengths by $Z_{49} $:
\begin{mathletters}
\begin{eqnarray}
{\tilde r} &\equiv& r/Z_{49} \\
{\tilde z} &\equiv& z/Z_\mathrm{49} \, \, ,
\end{eqnarray}
\end{mathletters}  
and all densities by $n_{49}$:
\begin{mathletters}
\begin{eqnarray}
\label{NondimensionalDensitya}
{\tilde n_0} &\equiv& n_0/n_{49} \\
{\tilde n} &\equiv& n/n_\mathrm{49} \, \, .
\label{NondimensionalDensityb}
\end{eqnarray}
\end{mathletters}
Since the ionized flow is transonic, we normalize all velocities to the ionized sound speed $a_I$, which we fix at 10 km/s:
\begin{equation}
\tilde{u} = u/a_I
\label{FiducialVelocity} \, \, .
\end{equation}

We introduce a nondimensional expression for the force per mass due to radiation pressure:
\begin{equation}
\tilde{f}_\mathrm{rad} \equiv f_\mathrm{rad} \, \left( \frac{a_I^2}{Z_{49}} \, \right )^{-1} \, \, .
\label{NonDimensionalFrad}
\end{equation}
This nondimensional quantity is the radiative force relative to that from thermal pressure. Then, using equation (\ref{FradDefinition}) for $f_{\rm rad}$, we have
\begin{equation}
\tilde{f}_\mathrm{rad} = \langle \cos \theta \rangle \left(\frac{\epsilon/c}{m_H \, a_I}\right) \left\{ \frac{Z_{49}/a_I}{[(n/2) \, \alpha_B]^{-1}}\right \} \, \, .
\label{FradFactored}
\end{equation}
The second factor on the right is the ratio of the momentum of an ionizing photon to the thermal momentum of a gas particle. The third factor is the ratio of the sound crossing time in the flow to the local photon recombination time.

\begin{figure}
\plotone{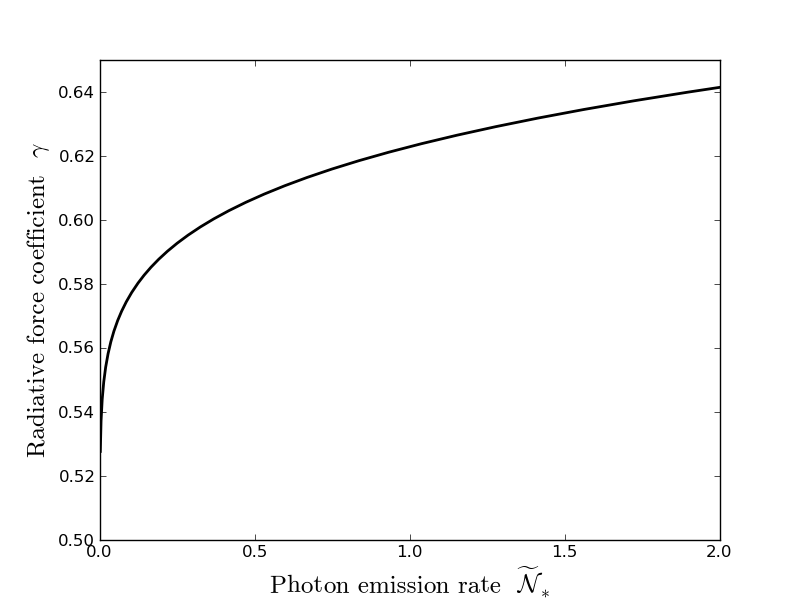}
\caption{The variation of the radiative force coefficient $\gamma$ defined in equation (\ref{GammaDefinition}) with $\tilde{{\cal N}}_\ast$, the star's ionizing photon emission rate, normalized to $10^{49}$ s$^{-1}$.}
\label{GammaVaryN}
\end{figure}

Since we are fixing the ionized sound speed, the only quantities that vary with $z$ in equation (\ref{FradFactored}) are $n$ and  $\langle \cos \theta \rangle$.\footnote{Strictly speaking, $\epsilon$ also varies with position. Higher-energy photons have a lower photoionization cross section, and thus travel farther from the star before they are absorbed, leading to a gradual hardening of the radiation with increasing distance. In our wavelength-independent analysis, we ignore this effect.} Thus, we are motivated to write
\begin{equation}
\tilde{f}_\mathrm{rad} = \gamma \, \tilde{n} \, \langle \cos \theta \rangle \,  \, ,
\label{GammaIntroduction}
\end{equation}
where $\gamma$ is also nondimensional. After setting the righthand sides of equations (\ref{FradFactored}) and (\ref{GammaIntroduction}) equal to each other, and making use of equations (\ref{FiducialDensity}), (\ref{FiducialLength}) and (\ref{NondimensionalDensityb}), we find that $\gamma$ can be expressed as 
\begin{equation}
\gamma \equiv \frac{3}{2} \, \left( \frac{L_{49}}{c}\right)\left(\frac{a_\mathrm{cl}^4}{G}\right)^{-1} \left(\frac{\epsilon}{\epsilon_{49}}\right) \, \, .
\label{GammaDefinition}
\end{equation}
Here $L_{49} \equiv {\cal N}_{49} \, \epsilon_{49}$ is the luminosity (in erg s$^{-1}$) of a star of spectral type O7.5, and $\epsilon_{49}$ is the mean energy of ionizing photons emitted from such a star. We show in Appendix \ref{GammaNRelation} that $ \epsilon $, and hence $\gamma$, is a weak function of $\tilde{{\cal N}}_\ast$. The function $\gamma(\tilde{{\cal N}}_\ast)$ is plotted in Figure \ref{GammaVaryN}. In summary the three free parameters that we vary between calculations are $\tilde{n}_0$, $\tilde{{\cal N}}_\ast$, and the star's displacement from the midplane, $\zeta \equiv H_\ast / Z_{49}$. 

We now summarize our nondimensional equations. After dropping the tilde notation for the rest of this section, the decoupled equations of motion are 
\begin{equation}
\frac{du}{dz} = \left(\frac{1}{1 - u^2}\right)\left[-2 \,u \, \frac{1}{R} \, \frac{dR}{dz} (1  + v_\mathrm{inj}^2) + 2 \, v_\mathrm{inj} \frac{1}{R}\,\sqrt{1 + \left(\frac{dR}{dz}\right)^2}\, (1 + u^2)  + u \frac{d\Phi_{\rm cl}}{dz}  - u \, n \, \gamma \langle \cos \theta \rangle \right]
\label{DudzNondimensional}
\end{equation}
\begin{equation}
\frac{dn}{dz} = \left(\frac{1}{1 - u^2}\right)\left[2 \, n \, \frac{1}{R} \, \frac{dR}{dz} (u^2 + v_\mathrm{inj}^2)  - 4 \, n \, \frac{1}{R}\, \sqrt{1 + \left(\frac{dR}{dz}\right)^2} \,u \,v_\mathrm{inj} - n \frac{d\Phi_{\rm cl}}{dz}  + n^2 \,  \gamma \langle \cos \theta \rangle \right] \, \, .
\label{DndzNondimensional}
\end{equation}
The expression for $\langle \cos \theta \rangle$ is still given by equation (\ref{AverageCosine}) if we use the appropriate nondimensional lengths. From equation (\ref{VinjPrime}), the injection velocity is
\begin{equation}
v_\mathrm{inj} = \sqrt{\frac{16 n_\mathrm{cl}^2/\beta - 4 n \,  n_\mathrm{cl}}{4 n \, n_\mathrm{cl} - n^2}} \, \, ,
\label{InjectionVelocityNondimensional}
\end{equation}
where $\beta$ is fixed at $(10/2)^2 = 25$.

The nondimensional cloud density and gravitational potential are
\begin{equation}
n_{\rm cl} \,=\,{n_0} \ {\rm sech}^2\, \left[ (z\,+\,\zeta) \sqrt{  \frac{ {4 \, n_0}}{\beta} }\right] \, \, ,
\label{SlabDensityNondimensional}
\end{equation}
\begin{equation}
\Phi_{\rm cl} = -\frac{1}{\beta}\ln {\rm sech}^2 \,\left[(z\,+\zeta)\sqrt{ \frac{4 \, n_0}{\beta}}\right] \,\, .
\end{equation}
Finally, the simplified ray tracing equation (\ref{RayTracingApproximate}) becomes  
\begin{equation}
\frac{1}{3} {{\cal N}_\ast}  \,=\, \int_0^{r_f(\theta)}\!n^2(z)\,r^2\,dr \,\,.
\label{RayTracingNondimensional}
\end{equation}

\subsection{Numerical method}
\label{SolutionStrategy}
The shape of the ionization front can be obtained through equation (\ref{RayTracingNondimensional}), but only \emph{after} the density $n (z)$ is established. Since we do not know this density a priori, we begin with a guessed function. We then trace out the ionization front, and thus establish $R(z)$. We next calculate both $n(z)$ and $u (z)$ by integrating the coupled equations (\ref{DudzNondimensional}) and (\ref{DndzNondimensional}). This procedure yields a new density distribution $n (z)$ which we use to retrace the locus of the ionization front, again using equation (\ref{RayTracingNondimensional}). The process is repeated until convergence is reached.

To integrate equations (\ref{DudzNondimensional}) and (\ref{DndzNondimensional}), we must specify values of $n$ and $u$ at the base of the flow, where \hbox{$z = R = 0$}. These two initial values are not independent. We show in Appendix \ref{BaseBoundaryConditions} that $u(-H_b) = v_{\rm inj}(-H_b)$, where $ v_{\rm inj}(-H_b)$ is the injection speed at the base, as found from equation (\ref{InjectionVelocityNondimensional}). Since we know $n_{\rm cl}(-H_b)$ from equation (\ref{SlabDensityNondimensional}), $v_{\rm inj}(-H_b)$ is solely a function of $n(-H_b)$, and thus $u(-H_b)$ is too. In practice, therefore, we need only guess $n(-H_b)$. Also derived in Appendix \ref{BaseBoundaryConditions} are expressions for $du/dz$ and $dn/dz$ at the base, where the righthand sides of equations (\ref{DudzNondimensional}) and (\ref{DndzNondimensional}) have divergent terms. 

What is the correct value of $n(-H_b)$? The righthand sides of both equations (\ref{DudzNondimensional}) and (\ref{DndzNondimensional}) have prefactors that diverge when $u=1$. Thus, crossing the sonic point requires special care. For an arbitrarily guessed $n(-H_b)$, $u(z)$ either diverges upward or declines toward zero as the sonic point is approached. This behavior is a generic feature of wind problems, and a bifurcation procedure is often employed to pinpoint the physical flow. We use the method of ``shooting and splitting'' (Firnett 1974). Here, we repeatedly guess $n(-H_b)$, and successive densities farther downstream, until the velocity profile is established to within a preset tolerance. Specifically, iterations continue until the range of this accurate profile include $u$-values sufficiently close to 1, typically 0.98 or 0.99. 
  
To jump over the sonic point, we use the current values of $du/dz$ and $dn/dz$ to perform a single first-order Euler integration step, typically with a $z$-increment of 0.01 -- 0.03, depending on the values of the derivatives. Once we are downstream from the sonic point, we revert to direct integration of equations  (\ref{DudzNondimensional}) and (\ref{DndzNondimensional}).

A key feature of the flows we generate is that the density near the base is low enough that the ionized gas is \emph{underpressured} with respect to the neutral cloud. The pressure drop causes neutral gas to be drawn into the flow and thereby replenish it. Driven by a combination of thermal and radiative forces, the flow accelerates. Thus, its density falls, but the decline is mitigated by the continual influx of fresh gas. On the other hand, the \emph{cloud} density always falls sharply (see equation \ref{SlabDensityNondimensional}). Eventually, $n_{\rm cl}(z)$ reaches the value $\beta n(z) / 4$, at which point the ionized and neutral gas have equal pressures. According to equation (\ref{InjectionVelocityNondimensional}), no more neutral gas is drawn into the flow beyond this point. In reality, the flow diverges laterally and its density also falls steeply. We do not follow this spreading process, but end each calculation at the point where the pressures cross over.  

\section{Results}
\label{Results}

\subsection{Fiducial model characteristics}

\begin{figure}
\plotone{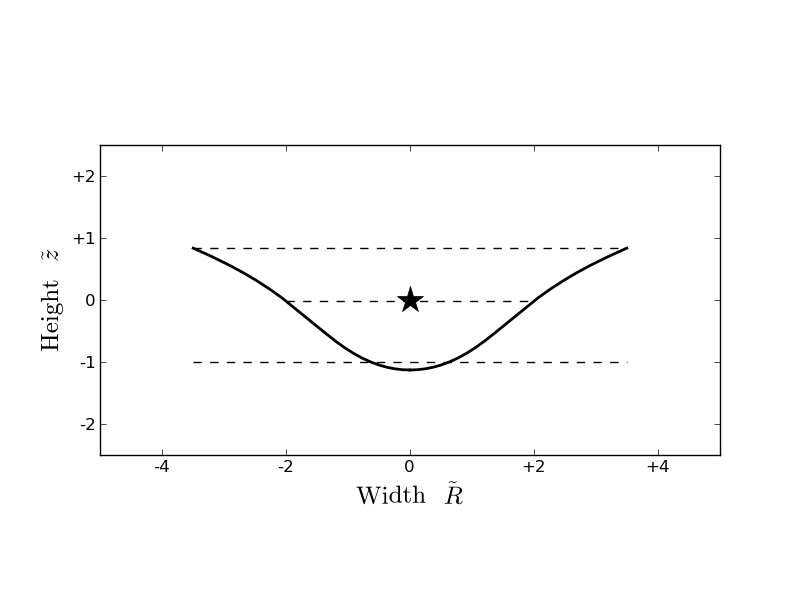}
\caption{Converged ionization front shape for our fiducial UCHII region, with the star's position indicated. In this and succeeding figures, all physical variables are displayed nondimensionally. The lowest dashed line represent the $z$-position of the cloud midplane. The middle dashed line represents the $z$-position of the sonic point. Finally, the upper dashed line is the endpoint of our solution, where the pressures of the ionized and neutral gas cross over.}
\label{FiducialIfront}
\end{figure}

For our fiducial model, we set the three nondimensional parameters to: $\tilde{n}_0 =\beta/4$; $\tilde{\cal N}_\ast = \zeta = 1$. The value of $\tilde{n}_0$ is chosen so that the cloud midplane is in pressure balance when the flow density $\tilde{n}$ is unity. Dimensionally, the midplane density is $8.8 \times 10^4$ cm$^{-3}$, the star's photon emission rate is ${\cal N}_{\ast} = 1 \times 10^{49}$ s$^{-1}$, and the star is displaced from the midplane by $H_\ast = 0.18$ pc. Figure \ref{FiducialIfront} shows the converged shape of the ionization front for this case. The rapid flaring of the base essentially reproduces the typical cometary shapes observed. Of course, a more precise comparison is between the predicted and observed emission measures. Here, too, the qualitative agreement is good, as we will later demonstrate.

The lowest dashed line in Figure \ref{FiducialIfront} represents the cloud midplane. Note that the base of the ionization front lies slightly below it. We have also displayed, as the middle dashed line, the sonic transition. In this particular model, the flow speed reaches $a_I$ close to the $z$-position of the star. This near match does not hold throughout most of parameter space. For example, lowering the stellar emission rate ${\cal N_\ast}$ moves the sonic transition farther above the star. Finally, the uppermost dashed line marks the height where the internal pressure of the flow overtakes that of the parent molecular cloud. In this case, the crossover point is about the same distance from the star as the base of the flow. As explained previously, we end our calculation at the crossover point, and do not attempt to track the complex dynamics of the flow as it continues to spread laterally and deposit its momentum into the lower-density surrounding gas. In any event, there are fewer observational constraints on this more diffuse flow.

\begin{figure}
\plotone{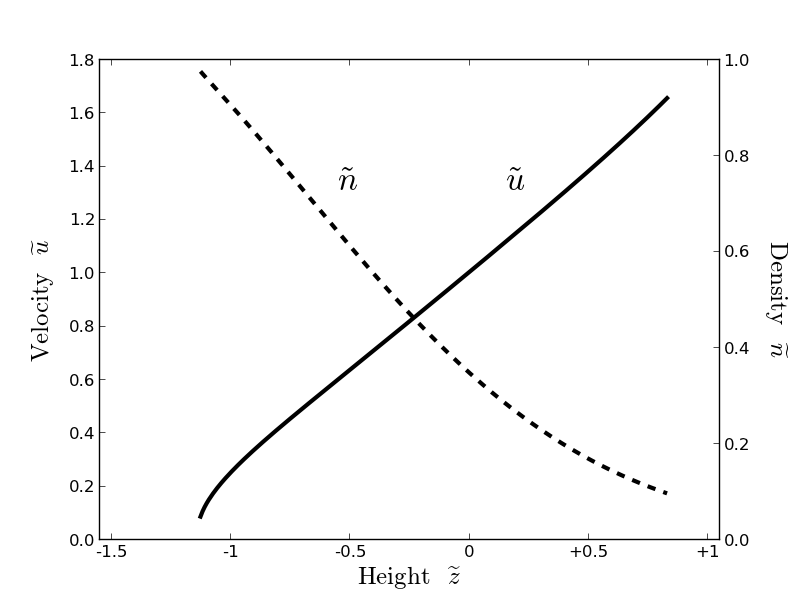}
\caption{Ionized gas density and velocity as a function of $z$-position for our fiducial model.}
\label{FiducialDensityAndVelocity}
\end{figure}

Figure \ref{FiducialDensityAndVelocity} shows the run of the density and velocity of the ionized gas with $z$-position. The velocity displays a smooth sonic transition, and a nearly constant acceleration throughout. The velocity at the base is very close to zero, but this is not the case for other parameter choices, as we will describe in the following section. As the velocity rises, the density falls, creating a pressure gradient that works to accelerate the flow.

To analyze more quantitatively the flow dynamics, it is helpful to gauge the relative contributions of the various terms contributing to the overall momentum balance. The nondimensional version of equation (\ref{MomentumConservation}) is 
\begin{equation}
u \, \frac{du}{dz} = - \frac{dn}{dz}  - \frac{d\Phi_{\rm cl}}{dz} + f_\mathrm{rad} + \frac{2}{R} \, \frac{dR}{dz} \, v_\mathrm{inj}^2  - \frac{2}{ R} \, \sqrt{1 + \left(\frac{dR}{dz}\right)^2} \, v_\mathrm{inj} \, u \, .
\label{NondimensionalMomentumConservation}
\end{equation}
Recall that the fourth and fifth righthand terms represent, respectively, the ram pressure from the injected gas and the retarding effect of mass loading.

\begin{figure}
\plotone{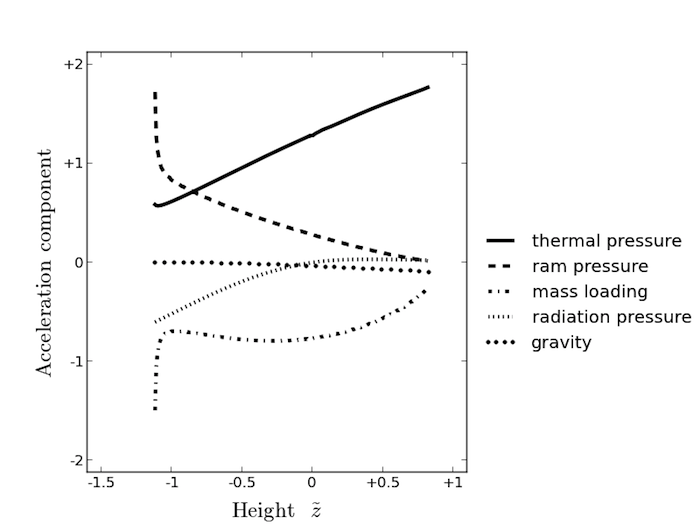}
\caption{The magnitude of each of the terms in the momentum equation (\ref{NondimensionalMomentumConservation}) as a function of $z$-position. The thermal pressure gradient, along with the ram pressure from injection of ionized gas, act to accelerate the flow, while mass loading and the gravity of the parent cloud decelerate it. Radiation pressure mostly acts to decelerate the flow, but for $z>0$ it makes a small contribution toward accelerating it.}
\label{FiducialMomentumTerms}
\end{figure}

Figure \ref{FiducialMomentumTerms} plots all five righthand terms as a function of $z$-position in the flow. The injected ram pressure and mass loading dominate at first, and nearly balance one another. Soon, however, the thermal pressure gradient takes over and remains dominant thereafter. The radiation force at first retards the flow, and is a minor contributor to its acceleration above the star's position. Finally, it is noteworthy that the gravitational force is negligible for most of the flow.

\subsection{Parameter Variation}

\begin{figure}
\plotone{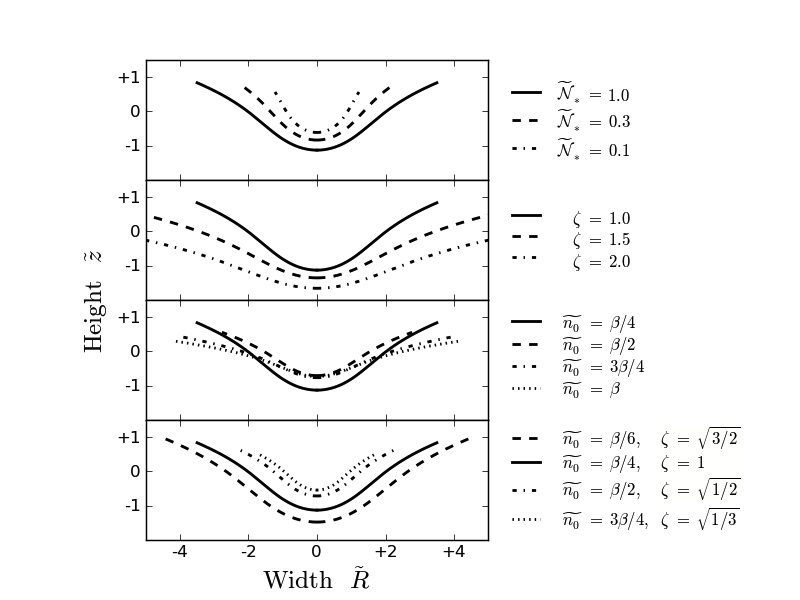}
\caption{Effect of parameter variation on the shape of the ionization front. Within each panel, the star is located at (0,0). In this and subsequent graphs, the fiducial model is indicated by a solid black line. For the entire study, $\beta$ is set to 25. } 
\label{IfrontVaryEverything}
\end{figure}

The top panel of Figure \ref{IfrontVaryEverything} shows how the shape of the converged ionization front changes with the stellar photon emission rate ${\cal N}_\ast$, while the cloud midplane density $n_0$ and stellar displacement from the midplane $\zeta$ are held fixed at their fiducial values. In this case the variation is predictable. A higher photon emission rate allows the ionizing photons to penetrate farther into the neutral cloud, in a manner similar to how the Str\"omgren radius of a spherical HII region increases with emission rate. A larger luminosity also leads to more flaring of the ionization front at large values of $z$.

We did not consider systems with photon emission rate ${\cal N}_\ast < 0.1$ because, at those low luminosities, the fraction of the star's luminosity that is composed of ionizing photons drops rapidly. We also found that, given our fiducial values of $n_0$ and $\zeta$, we could not achieve transonic solutions with ${\cal N}_\ast \gtrsim 1$. As can be seen from the plot of velocity, $u$ is already quite close to $0$ for ${\cal N}_\ast = 1$. Attempting to raise the photon emission rate any higher without changing the other parameters creates such a large ionized density near the base that the flow is no longer underpressured with respect to the cloud. It is possible to create flows with higher values of ${\cal N}_\ast$ if, for example, $\zeta$ is increased at the same time.

\begin{figure}
\plotone{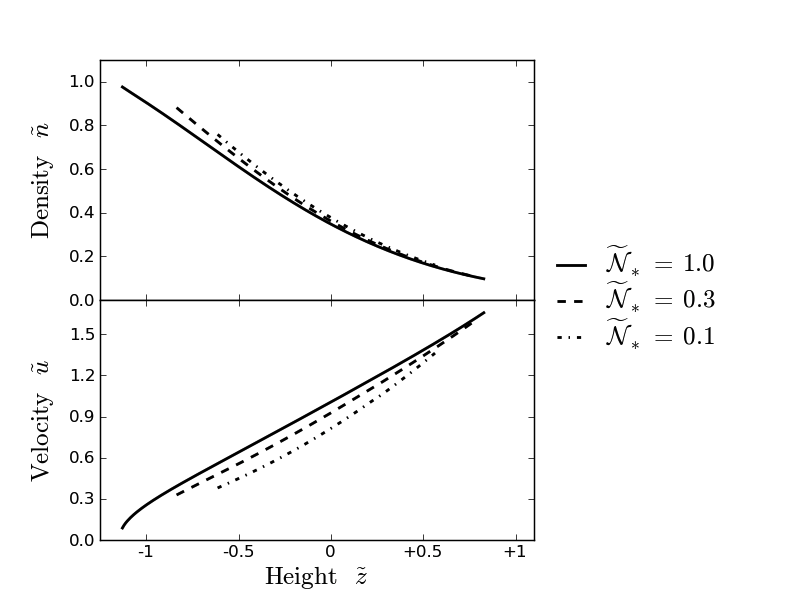}
\caption{Density and velocity profiles for various ionizing photon emission rates. The stellar displacement $\zeta$ and the midplane cloud density $n_0$ are held fixed at their fiducial values of 1 and $\beta/4$, respectively. }
\label{VelocityVaryLuminosity}
\end{figure}

Generally, we find that the density profile of the ionized flow mimics that of the neutral cloud. Figure \ref{VelocityVaryLuminosity} shows how the density and velocity profiles vary when the stellar luminosity is changed. The flow structure remains strikingly similar as the luminosity is varied, changing primarily in spatial extent, because a higher luminosity allows ionizing radiation to penetrate farther into the cloud. The fact that the density and velocity profiles do not show other significant variations with luminosity reinforces the conclusion that it is the density structure of the neutral cloud that sets the spatial variation in the ionized flow.

The second panel of Figure \ref{IfrontVaryEverything} shows how the shape of the converged ionization front changes with the stellar displacement $\zeta$. Here, the effect is similar to that of increasing ${\cal N}_\ast$. With higher $\zeta$, the ionizing photons penetrate a larger distance into the cloud, in this case because they encounter a lower density when the star is displaced farther from the cloud midplane. Note also the flaring at large $z$, which increases sharply for higher $\zeta$.

\begin{figure}
\plotone{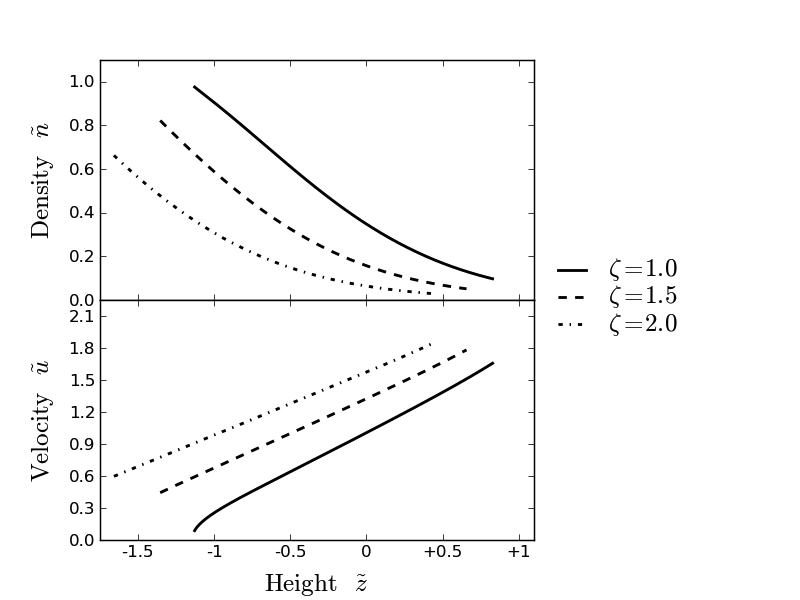}
\caption{Density and velocity profiles for various values of the stellar displacement $\zeta$. The photon emission rate ${\cal N_\ast}$ and the midplane density $n_0$ are held fixed at their fiducial values of 1 and $\beta/4$, respectively}
\label{VelocityVaryZeta}
\end{figure}

Figure \ref{VelocityVaryZeta} shows how the density and velocity profiles vary when $\zeta$ is changed. Again, the ionized gas density tracks that of the neutral cloud. For larger offsets, the flow begins in a less dense portion of the cloud, and the density of the ionized flow is smaller. Since the ionizing stellar photons penetrate farther into the cloud when $\zeta$ is larger, the flow begins at more negative values of $z$. The velocity profiles shift along the $z$-axis as $\zeta$ is varied, but the acceleration remains roughly the same. 

Attempting to lower $\zeta$ below a value of unity, while keeping $n_0$ and ${\cal N}_\ast$ fixed at their fiducial values, results in the same problem as attempting to increase ${\cal N}_\ast$ on its own. Solutions with lower values of $\zeta$ can be achieved only if $n_0$ is simultaneously increased or ${\cal N}_\ast$ is decreased. Increasing $\zeta$ alone also leads to a large amount of flaring of the ionization front. For $\zeta = 2$, the flaring at the location of pressure crossover is such that $R/H_b = 5.3$. As we shall see in section \ref{EmissionMeasureMaps}, this aspect ratio is large compared to observations.

The third panel of Figure \ref{IfrontVaryEverything} shows how the shape of the ionization front changes when $n_0$ is varied on its own. In this case, the variation is more complex. As can be seen from equations (\ref{SlabDensity}) and (\ref{ScaleHeight}), changing $n_0$ changes not only the overall magnitude of the density profile, but also the steepness of its falloff. When $n_0$ is increased from $\beta/4$ to $\beta/2$, the rise in the midplane density means that radiation cannot penetrate as far, and the base of the ionized flow moves closer to the star. However, as $n_0$ is increased further to $3 \beta / 4$, the steeper falloff of the density comes into play, and the distance between the star and the base of the flow remains nearly constant.

We may effectively factor out the increasing cloud scale height if, instead of increasing $n_0$ on its own, we simultaneously \emph{decrease} $\zeta$ so that $H_\ast / H_{\rm cl}$ remains equal to unity. This result of this exercise is shown in the bottom panel of Figure \ref{IfrontVaryEverything}. As $n_0$ increases and $\zeta$ decreases, the shape of the flow remains strikingly similar, changing primarily in spatial extent. In all cases, $H_b$/$H_{\rm cl}$ remains close to unity, ranging from $1.21$ when $n_0 = \beta/6$ to $0.94$ when $n_0 = 3 \beta / 4$.

\begin{figure}
\plotone{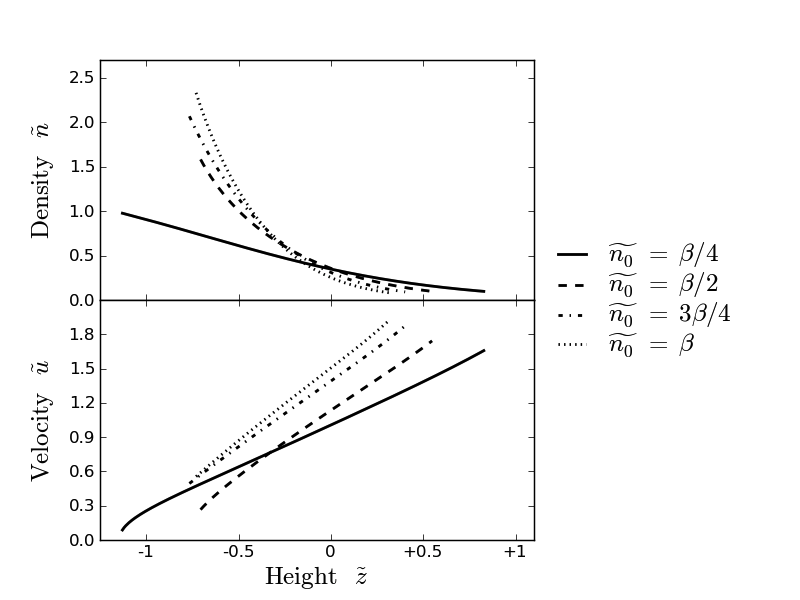}
\caption{Density and velocity profiles for various midplane densities of the neutral cloud, $n_0$. The nondimensional ionizing photon emission rate ${\cal N_\ast}$ and stellar offset from the midplane $\zeta$ are held fixed at their fiducial values of 1.}
\label{VelocityVaryN0}
\end{figure}

Figure \ref{VelocityVaryN0} shows how the density and velocity profiles respond to changes in $n_0$. For a denser neutral cloud, the ionized flow also begins at a larger density. At the same time, the density of both the neutral and ionized gas drop more rapidly for a larger $n_0$, and the $z$-extent of the flow diminishes. Finally, since the ionized gas is primarily accelerated via pressure gradients, a steeper decline in density also corresponds to a more rapid acceleration. 

\begin{figure}
\plotone{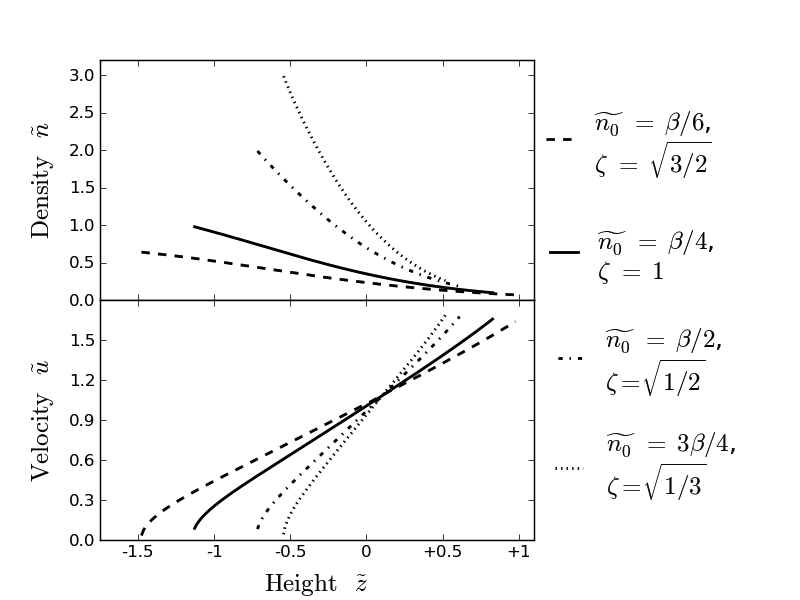}
\caption{Density and velocity profiles for various flow models. The midplane density $n_0$ and stellar displacement $\zeta$ are varied simultaneously so as to keep the ratio $H_\ast$/$H_{\rm cl} = 1$.}
\label{VelocityVaryN0AndZeta}
\end{figure}

We may again factor out the effect of scale height by varying $n_0$ and $\zeta$ simultaneously in the manner described previously. The resulting density and velocity profiles are shown in Figure \ref{VelocityVaryN0AndZeta}. In this case the density at the base of the ionized flow again increases with rising $n_0$, but now in a more systematic manner. In fact, this base density remains almost exactly equal to $4 n_0 /\beta$, reflecting the fact that the flows are beginning at pressures very close to the neutral cloud pressure. This near-pressure equality also accounts for the fact that the velocities begin near zero in all of these flows. The velocity profiles are remarkable for the fact that they share an anchor point near the stellar position at $z=0$. Their $z$-length scales correlate tightly with the changing cloud scale height.

\subsection{Mass, momentum and energy transport}
We next consider the total rate at which the neutral cloud adds mass to the ionized flow. Dimensionally, this rate, from the base to any height $z$, is 
\begin{equation}
\dot{M}(z) \equiv 2 \pi \, \int_{-H_b}^z {\rho \, v_\mathrm{inj} \,R \, \sqrt{1 + \left(\frac{dR}{dz}\right)^2} dz }\, \, .
\label{MdotDefinition1}
\end{equation}

\begin{figure}
\plotone{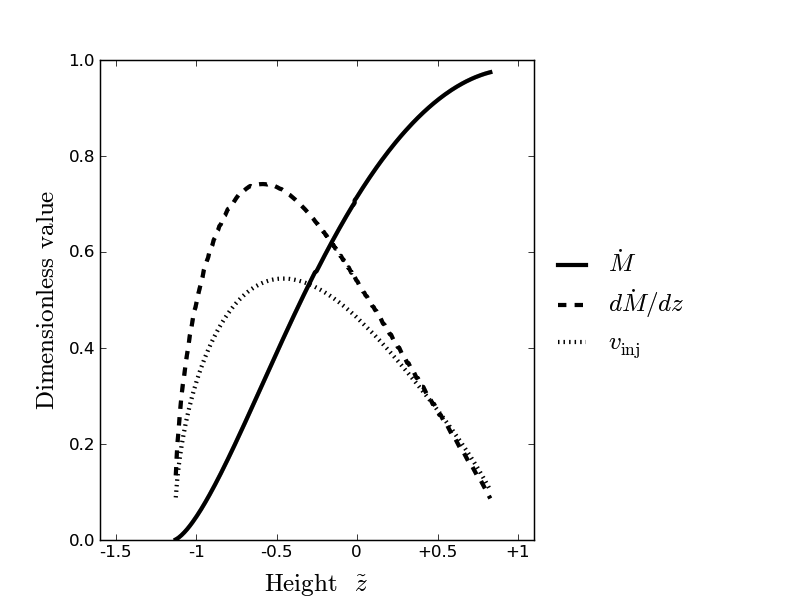}
\caption{Nondimensional mass injection rate and related quantities as functions of $z$. The mass injection rate $\dot{M}(z)$ has been normalized to $\dot{M}_0 = 3.7 \times 10^{-4} \Msun$ yr$^{-1}$ (see equation \ref{FiducialMdot}). Its derivative $d\dot{M}/dz$ is normalized to $\dot{M}_0/ Z_{49} = 2 \times 10^{-3} \Msun$ yr$^{-1}$ pc$^{-1}$. Finally, the injection speed $v_{\rm inj}$ has been normalized to the ionized sound speed, assumed to be 10 km s$^{-1}$.}
\label{MassInjectionTerms}
\end{figure}

Figure \ref{MassInjectionTerms} displays $\dot{M}(z)$ in our fiducial model. The cross-sectional area of the flow starts at zero and monotonically increases with $z$. The mass injection rate also climbs from zero, but levels off at the pressure crossover point, where no new mass is being added. We have also plotted the injection speed, $v_{\rm inj}(z)$. This starts out relatively small, peaks at about half the ionized sound speed at $z \approx -0.5$, and then eventually falls to zero. As the figure also shows, $d\dot{M}/dz$ attains its maximum close to where $v_{\rm inj}$ peaks. 

We stop our calculation at the pressure crossover location $z_f$, where neutral material ceases to be drawn into the ionized flow. In steady state, the rate at which ionized mass crosses $z_f$ should equal the total rate of mass injection up to this point. That is, the mass \emph{outflow} rate in ionized gas should be 
\begin{equation}
\dot{M}_{\rm out} = \dot{M}(z_f) 
\end{equation}
We may also calculate $\dot{M}_{\rm out}$ using the local dimensional relation
\begin{equation}
\dot{M}_{\rm out} = \pi \, R_f^2 \, \rho_f \, u_f \, \, .
\label{MdotDefinition2}
\end{equation}
We find that we obtain the same mass outflow rate using these two methods to within a precision of 0.1 percent, providing a useful check on mass conservation.

Dimensionally, the mass outflow rate is 
\begin{align}
\dot{M}_{\rm out} & = 2\pi \left(\frac{m_\mathrm{H}}{2}\right) \,  n_{49}  \, Z_{49} ^2 \, a_I \, {\cal I}_f \\ & = \frac{a_{\rm cl}^4}{G a_I} \, {\cal I}_f \\ & = 3.7 \times 10^{-4} \, M_{\odot} \, \, {\rm yr}^{-1} \, {\cal I}_f \, \, ,
\label{FiducialMdot}
\end{align}
where the nondimensional quantity ${\cal I}_f$ is 
\begin{equation}
{\cal I}_f \equiv \int_0^{z_f}{{n}\, {v_\mathrm{inj}} \, {R} \sqrt{1 + \left ( \frac{dR}{dz}\right)^2 } \, d{z}} \, \, .
\end{equation}
The dependence of $\dot{M}$ on our three parameters ${\cal N}_\ast$, $n_0$, and $\zeta$ is entirely contained within ${\cal I}_f$.

We also define a momentum outflow rate $\dot{p}_{\rm out} \equiv \dot{M}_{\rm out} u_f$. This quantity may be written 
\begin{align}
\dot{p}_{\rm out} & = 2\pi \left(\frac{m_\mathrm{H}}{2}\right) \,  n_{49}  \, Z_{49} ^2 \, a_I \, u_f\, {\cal I}_f \\ 
& =\left(\frac{a_{\rm cl}^4}{G}\right) \left(\frac{u_f}{a_I} \right) \, {\cal I}_f \\ 
& = 2.4 \times 10^{28} \, {\rm dyne} \left(\frac{u_f}{a_I}\right) {\cal I}_f \, \, .
\end{align}
Finally, the dimensional kinetic energy outflow rate, $\dot{E}_{\rm out} \equiv 1/2 \dot{M}_{\rm out} u_f^2$, is 
\begin{align}
\dot{E}_{\rm out} & = \left( \frac{1}{2} \right) \, 2\pi \, \left(\frac{m_\mathrm{H}}{2}\right) \,  n_{49}  \, Z_{49} ^2 \, a_I \, u_f^2\, {\cal I}_f \\ 
& =\left(\frac{a_{\rm cl}^4 a_I }{2 G}\right) \left(\frac{u_f}{a_I} \right)^2 \, {\cal I}_f \\ 
& = 1.2 \times 10^{34} \, {\rm erg} \, \, {\rm s}^{-1} \, \, \left(\frac{u_f}{a_I}\right)^2 {\cal I}_f \, \, .
\end{align}

Table \ref{MassLossTable} displays the values of $\dot{M}_{\rm out}$, $\dot{p}_{\rm out}$ and $\dot{E}_{\rm out}$ for a variety of parameter choices. The mass loss rate is far more sensitive to changing parameters than is $u_f$, so the trends in momentum and energy transport can be explained almost entirely by the trends in $\dot{M}_{\rm out}$. Moreover, as equation (\ref{MdotDefinition2}) demonstrates, $\dot{M}_{\rm out}$ is principally affected by changes in the cross-sectional area and density at $z_f$. 

As before, we first examine the effect of varying the photon emission rate. With increasing ${\cal N}_\ast$, $R_f$ increases as well, driving up $\dot{M}_{\rm out}$. On the other hand, spreading of the flow results in a slight decline of the density $n_f$. This latter effect weakens the sensitivity of $\dot{M}_{\rm out}$ to ${\cal N}_\ast$. As the dimensional ${\cal N}_\ast$ rises from $10^{48}$ to $10^{49}$ s$^{-1}$, $\dot{M}_{\rm out}$ scales as ${{\cal N}_\ast}^{0.78}$. 

Similar trends appear when we increase the stellar offset $\zeta$, leaving other parameters fixed (see second level of Table 1). With larger $\zeta$, the ionization flow flares out. At the same time, the flow begins in a less dense portion of the cloud, so that $n_f$ falls. Nevertheless, the net effect is for $\dot{M}_{\rm out}$ to increase. In the range $1 < \zeta < 2$, we find that $\dot{M}_{\rm out}$ scales as $\zeta^{1.2}$.

Increasing $n_0$ on its own leads to complicated behavior similar to that we encountered previously. As $n_0$ rises from $\beta/4$ to $\beta/2$, $\dot{M}_{\rm out}$ first falls because the ionization front shrinks in size (see third panel of Figure \ref{IfrontVaryEverything}). However, as $n_0$ further increases from $\beta/2$ to $\beta$, the cloud scale height continues to shrink. The rapidly declining cloud density at any fixed $z$ causes the ionized outflow to broaden, and $\dot{M}_{\rm out}$ rises.

Finally, we may vary $n_0$ and $\zeta$ simultaneously, so as to keep $H_\ast/H_{\rm cloud} = 1$. The fourth panel of Figure \ref{IfrontVaryEverything} shows that the ionization front retains its shape but shrinks in scale. As Figure \ref{VelocityVaryN0AndZeta} shows, increasing $n_0$ under these circumstances raises the ionized density at any $z$. Consequently, the stellar radiation cannot penetrate as far. In particular, $R_f$ becomes smaller at the pressure crossover point. Table 1 verifies that this shrinking of the cross-sectional area causes $\dot{M}_{\rm out}$ to decline. Quantitatively, we find $\dot{M}_{\rm out} \propto n_0^{-0.37}$.

The fact that the mass outflow rate generally decreases with increasing cloud density is significant, and calls for a more basic physical explanation. The lateral size $R$ of the outflow is about that of a pressure-bounded HII region. From equation (\ref{StromgrenRadius}), we have $R \propto n_I^{-2/3}$, where $n_I$ is a representative ionized density in the flow. From equation (\ref{MdotDefinition2}), the mass loss rate $\dot{M}_{\rm out}$ is proportional to $R^2 \, n_I$. Together these two relations imply $\dot{M}_{\rm out} \propto n_I^{-1/3}$. Finally, if $n_I$ is proportional to the peak neutral density $n_0$, as follows from the condition of pressure equilibrium, then we have $\dot{M}_{\rm out} \propto n_0^{-1/3}$, which is close to our numerical result.

This argument also helps to explain why the mass loss rates that we obtain are smaller than those of past champagne flow calculations. For example, \citet{Bodenheimer1979} found that, for a star with $ {\cal N}_\ast = 7.6 \times 10^{48}$ s$^{-1}$ embedded in a slablike molecular cloud of uniform number density $10^3$ cm$^{-3}$, the mass loss rate is $2 \times 10^{-3}$ $M_\odot$ yr$^{-1}$. For the much denser molecular cloud in our fiducial model, extrapolation of the \citet{Bodenheimer1979} rate using $\dot{M}_{\rm out} \propto n_0^{-1/3}$ yields $\dot{M}_{\rm out} = 4 \times 10^{-4}$ $M_\odot$ yr$^{-1}$, quite close to our calculated result. It should be kept in mind that this comparison is only meant as a consistency check, given the many differences in detail of the two calculations.

We may combine our scaling results for the mass outflow rate in order to show explicitly its dependence on ${\cal N}_\ast$ and $n_0$. We have

\begin{equation}
\dot{M}_{\rm out} = 3.5 \times 10^{-4} \, M_{\odot} \, \, {\rm yr}^{-1} \left(\frac{{\cal N}_{\ast}}{10^{49} \, \, {\rm s}^{-1}} \right)^{0.78} \left(\frac{n_0}{10^{5} \, \, {\rm cm}^{-3}} \right)^{-0.37} \, \, ,
\label{MdotScalingSummary}
\end{equation}

More precisely, the two power-law indices are $0.775 \pm 0.05$ and $-0.369 \pm 0.004$, where we have included the standard errors from our linear least squares fit. In writing equation (\ref{MdotScalingSummary}) we have implicitly assumed that the stellar displacement $H_\ast$ varies with $n_0$ so that the former equals one cloud scale height. We also caution that all these numerical results are based on a slab model for the parent cloud. Bearing these caveats in mind, equation (\ref{MdotScalingSummary}) may prove useful for future modeling of massive star formation regions.

The ionized wind in an UCHII region represents a large increase in mass loss over what the driving star would achieve on its own. A bare, main-sequence O7.5 star with ${\cal N}_\ast = 1 \times 10^{49}$ s$^{-1}$ radiatively accelerates its own atmosphere to create a wind with $10^{-7}$ to $10^{-6}$ $M_\odot$ yr$^{-1}$ \citep{Mokiem2007, Puls2008}, two to three orders of magnitude below our fiducial value. However, one cannot completely ignore the dynamical effect of the stellar wind in an UCHII region, as we will discuss shortly. Even younger stars of the same luminosity drive bipolar outflows that have far \emph{greater} mass loss rates than we compute, typically $10^{-3}$ to $10^{-2}$ $M_\odot$ yr$^{-1}$ \citep{Churchwell2002}. While the exact mechanism behind these molecular outflows is uncertain, they might result from the entrainment of cloud gas by massive jets emanating from an accreting protostar \citep{Beuther2005}.

Over the inferred UCHII lifetime of $10^5$ yr, a star driving an outflow at the rate of \hbox{$4 \times 10^{-4} \, \, M_\odot \,\, \rm{yr}^{-1}$} disperses 40 $M_\odot$ of cloud mass. This figure is tiny compared with $10^4$ $M_\odot$, the typical mass of a high-density clump within an infrared dark cloud \citep{Hofner2000, Hoare2007}. Thus the UCHII region, which arises in the densest part of the cloud structure, represents an early stage in its clearing. Presumably, a \emph{compact} HII region, of typical size 0.1 to 0.3 pc, appears as the star begins to clear less dense material. We expect the outflow region to broaden and the mass loss rate to increase during this longer epoch. 

Our O7.5 star drives a wind with an associated momentum output of about \hbox{$L_\ast / c = 3.0 \times 10^{28}\, \, \rm{dyne}$} \citep{Vacca1996}. This figure is remarkably close to the $\dot{p}_{\rm out}$ of $4 \times 10^{28}$ dyne in our fiducial model. The numerical agreement is fortuitous, since the ionized outflow represents material drawn in from the external environment and accelerated by thermal pressure. The correspondence between these two rates reflects the numerical coincidence that $a_{cl}^4/G \sim L_\ast / c$ for an embedded O star. In any event, a more complete treatment of the flow in an UCHII region would also account for the additional forcing from the stellar wind (see, e.g., \citeauthor{Arthur2006} 2006 for a simulation that includes this effect). Note finally that the ionizing photons themselves impart momentum to the flow. The resulting force is $f_{\rm rad}$, which we introduced in Section \ref{RadPressureExplained}, and whose quantitative effect we discuss below.

The kinetic energy transport rate for our fiducial model is only $4 \times 10^{-5}$ times $L_\ast$, the bolometric luminosity of the star. Given that the flow is transonic, the thermal energy carried in the outflow is of comparable magnitude. The vast bulk of the stellar energy is lost to radiation that escapes from the HII region during recombination, through line emission from ionized metals, and continuum, free-free emission \citep{Osterbrock1989}. For a time span of $10^5$ yr, the total energy ejected in our fiducial model is a few times $10^{47}$ erg, a figure comparable to the turbulent kinetic energy in a molecular clump of mass $10^4$ M$_\odot$ and internal velocity dispersion of 2 km s$^{-1}$. This match broadly supports the contention of \citet{Matzner2002} that HII regions provide the ultimate energy source of turbulence in molecular clouds large enough to spawn massive stars.

\subsection{Role of radiation pressure}
\label{RadiationPressure}

One of the novel features in this analysis of UCHII regions is our inclusion of the momentum deposition by ionizing photons. We derived $f_{\rm{rad}}$, the radiative force per mass, in Section \ref{RadPressureExplained} (see eq. (\ref{FradDefinition})), and expressed it nondimensionally in equation (\ref{GammaIntroduction}). How significant is this term in the overall flow dynamics?

\begin{figure}
\plotone{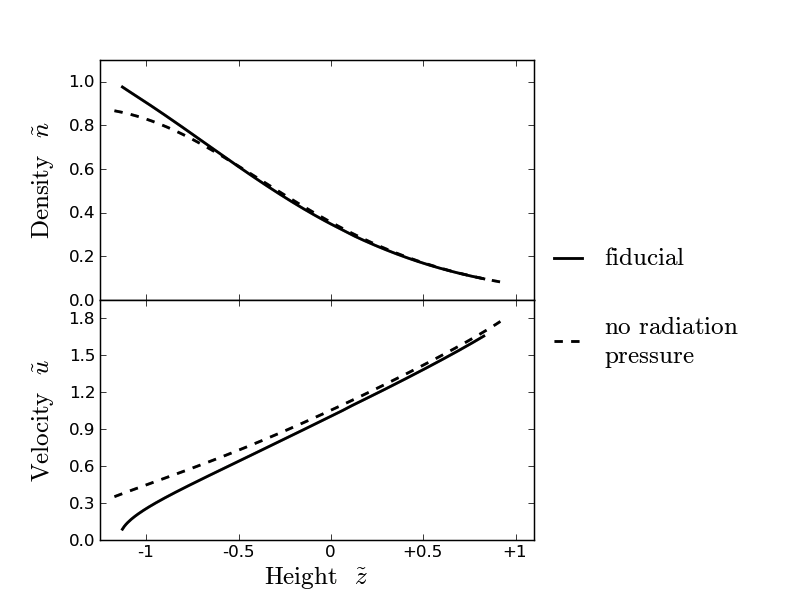}
\caption{Density and velocity profiles both for the fiducial
model ({\it solid curves}) and with radiation pressure omitted ({\it dashed curves}).}
\label{VelocityVaryRadpressure}
\end{figure}

To assess the role of $f_{\rm{rad}}$, we recalculated our fiducial model after artificially removing the force from the equations of motion,  (\ref{DudzNondimensional}) and (\ref{DndzNondimensional}). Figure \ref{VelocityVaryRadpressure} shows the result of this exercise. The dashed curve in the top panel is the altered density profile, $n(z)$, while the analogous curve in the bottom panel is the altered velocity, $u(z)$. 

Near the base, the stellar flux is directed oppositely to the ionized gas velocity. Thus, the radiation force {\it decelerates} the flow in this region. The bottom panel of the figure shows how the starting velocities are lower when radiation pressure is included. As a result of this force, gas piles up near the base, and therefore develops a {\it larger} thermal pressure gradient. The enhanced gradient drives the flow outward in spite of the retarding stellar photons. The density pileup is evident in the solid curve within the top panel of Figure \ref{VelocityVaryRadpressure}.   

The effect of radiation pressure diminishes at higher values of $z$. Recall that $f_{\rm{rad}}$ is proportional to the flow density $n$. Gas of higher density has a greater volumetric recombination rate, and thus absorbs more ionizing photons per time to maintain ionization balance. Since the density falls with distance $z$, so does the magnitude of the force.

A second contributing factor is the changing direction of photons emanating from the star. Near the base, the incoming flux is nearly all in the negative $z$-direction, so that $\langle \cos \theta \rangle$ is close to $-1$. This geometric term vanishes at the level of the star ($z = 0$), and then climbs back up toward $+1$. However, the rapidly falling density overwhelms the latter effect, and the magnitude of the force still declines (see the dotted curve in Figure \ref{FiducialMomentumTerms}).

Finally, the radiation force does not significantly alter the shape of the ionization front. In our fiducial model, the base of the flow is farther from the star when $f_{\rm rad}$ is omitted. However, the fractional change in this distance from the case with the force included is only 0.05.

\subsection{Bipolar outflows}
\label{BipolarOutflows}

The outflow topology sketched in Figure \ref{schematic} does not exist for arbitrary values of our parameters. If the star is situated too close to the densest portion of the cloud, or if the ionizing luminosity is too weak, then a transonic flow, steadily drawing in neutral material, cannot develop. At a sufficiently low luminosity or high neutral density, the HII region first undergoes pressure-driven expansion, but then remains trapped, i.e., density-bounded, on all sides.

Alternatively, with a much higher luminosity and/or lower cloud density, an outflow may erupt on \emph{both} sides of our model planar slab. Such a bipolar morphology is harder to achieve, in part because of the values of cloud density and stellar luminosity required, and in part because real clouds do not have a slab geometry, i.e., they are not infinite in lateral extent. Increasing the stellar luminosity in an originally monopolar flow usually just widens the ionization front, without creating another outflow lobe. It is therefore not surprising that out of the hundreds of UCHII regions that have been identified, only a handful have a bipolar morphology \citep{Garay1999, Churchwell2002}. 

Given their relative scarcity, we forgo a parameter study of this type of UCHII region, and focus instead on the simplest example. Here we place the star exactly at the cloud midplane, $\zeta = 0$. We set the cloud midplane density to $n_0 = \beta/4$, as in our fiducial, monopolar model. However, we find that ${\cal N}_\ast = 1$ does not lead to a bipolar flow. To be safe, we have raised ${\cal N}_\ast$ to 3. 

In order to generate this solution, we began with the fact that $u(0) = 0$, as demanded by symmetry. We guessed $n(0)$, the flow density at the midplane, and then used these two starting values as the initial conditions for integrating the coupled first-order equations (\ref{DudzNondimensional}) and (\ref{DndzNondimensional}). As before, we use equation (\ref{InjectionVelocityNondimensional}) to obtain the injection velocity. Since $R$ does not vanish at the base, the righthand sides of (\ref{DudzNondimensional}) and (\ref{DndzNondimensional}) are well-behaved at the start, and the integration is relatively straightforward. Using the method of shooting and splitting, we refine our guess for the starting ionized density until we approach the sonic point. We jump over this point in just the manner described in Section \ref{SolutionStrategy}.

\begin{figure}
\plotone{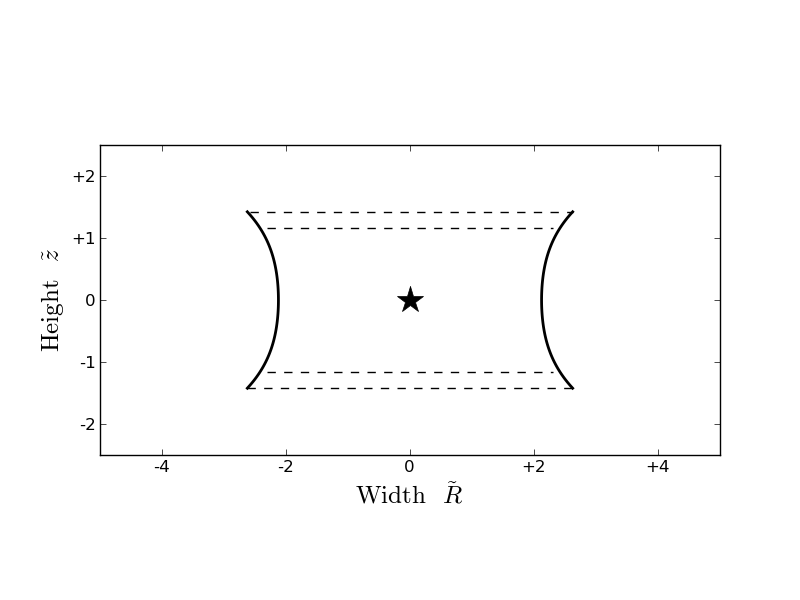}
\caption{The ionization front for a symmetric, bipolar transonic outflow. Here ${\cal N}_{49} = 3$, $n_0 = \beta/4$, and $\zeta = 0$. The two horizontal dashed lines closer to the star mark the sonic transitions, while the outer pair indicates the pressure crossover.}
\label{BipolarIfront}
\end{figure}

\begin{figure}
\plotone{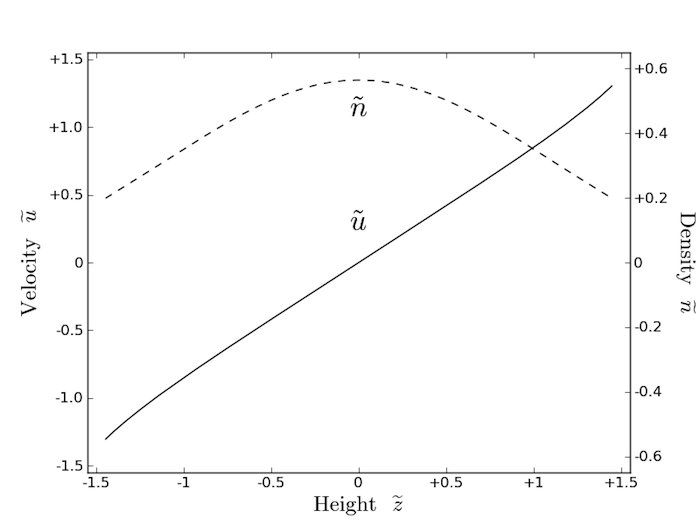}
\caption{Density and velocity profiles for the bipolar outflow shown in Figure \ref{BipolarIfront}.}
\label{BipolarDensityAndVelocity}
\end{figure}

Figure \ref{BipolarIfront} shows the symmetric ionization front for this model. The two sets of horizontal lines show the location of the sonic transitions and the pressure crossover points, respectively. We notice immediately how close these points are to one another (compare Fig. \ref{FiducialIfront}). This same feature is apparent in Figure \ref{BipolarDensityAndVelocity}, which displays the density and velocity profiles. The velocity, which is an odd function of the height $z$, reaches unity just before the curve ends at the pressure crossover. The ionized density $n(z)$ is symmetric about $z = 0$ and has a shape similar to that of the neutral cloud, peaking at the midplane. 

In this outflow, neutral gas with a pressure exceeding that of the ionized gas is drawn in laterally through the ionization front. The ionized gas flows away in a symmetric manner from its region of maximal density at the midplane. Radiation pressure never acts to decelerate the flow, as in the monopolar case. Rather, it contributes to the acceleration, consequently reducing the ionized density gradient. Nevertheless, the thermal pressure gradient remains the strongest driving force.

As mentioned previously, attempts to lower ${\cal N}_\ast$ closer to 1 resulted in failure to obtain a transonic solution. Specifically, the pressure crossover point was reached before the sonic transition. We found this result for any guess of the starting ionized density below $4 n_0 / \beta$, i.e. for ionized pressures at the midplane less than the corresponding cloud pressure. Our inability to find transonic solutions in this regime suggests that the true steady-state solution is not an outflow, but a trapped HII region in which the interior and cloud pressures match.

\subsection{Pseudo-cylindrical cloud}
\label{CylindricalCloud}

Up to this point we have taken our molecular cloud to be a planar slab, a geometry that is convenient mathematically, but not truly representative of the clouds found in nature. There exist no detailed studies of infrared dark cloud morphologies. In the realm of low-mass star formation, recent observations of Gould Belt clouds reveal that a tangled network of filaments creates the dense cores that in turn collapse to form stars \citep{Andre2010}. Thus, a more realistic model for our background cloud might be a cylinder in force balance between self-gravity and turbulent pressure. If the interior velocity dispersion is again spatially uniform, as we have assumed, then the density falls with a power law with distance from the central axis, as opposed to the much steeper, exponential falloff we have employed until now.

Our model is flexible enough that we can explore various functional forms for the cloud density profile. However, our cloud is still a one-dimensional slab, so that the density $\rho$ is a function of $z$, the distance from the midplane. We construct a “pseudo-cylindrical” cloud, which has the density profile of a self-gravitating, isothermal cylinder \citep{Ostriker1964}, but with the cylindrical radius $R$ replaced by $z$. In this model, the dimensional number density and gravitational potential are
\begin{equation}
n_{\rm cl} \,=\, n_0 \left [1 + \frac{1}{4}\left(\frac{z + H_\ast}{H_{\rm cl}}\right)^2 \right ] ^{-2} \,\, ,
\end{equation}
\begin{equation}
\Phi_{\rm cl} = -2 \, a_{\rm cl}^2 \, \ln \left [1 + \frac{1}{4}\left(\frac{z + H_\ast}{H_{\rm cl}}\right)^2 \right ] \, \, .
\end{equation}
Here $H_\ast$ is again the distance between the star and the cloud midplane, and $H_{cl}$ is the scale height in equation (\ref{ScaleHeight})\footnote{Our choice of formula for $H_{cl}$ is a factor of $\sqrt{2}$ larger than that used in \citet{Ostriker1964}.}. 

\begin{figure}
\plotone{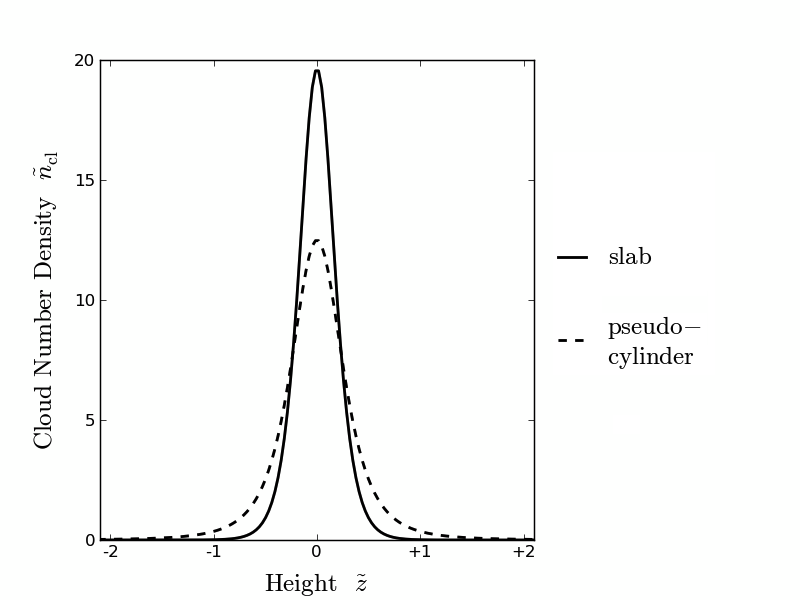}
\caption{Cloud density profiles for a slab and pseudo-cylindrical model of equal column density.}
\label{CloudDensity}
\end{figure}

Choosing ${\cal N}_\ast = 1$  and $\zeta = 1$, we could not achieve a transonic flow, if we also used the fiducial $n_0 = \beta/4$. In retrospect, this result could have been anticipated, since the pseudo-cylindrical model has a smaller column density, as measured from the midplane, than the slab model with the same $n_0$. We therefore utilized $n_0 = \beta/2$ in the pseudo-cylindrical case, for which we could indeed find a transonic solution. For comparison, we ran a slab model with $n_0 = \pi \beta/2$, which has the same column density as the pseudo-cylinder. The two density profiles, both non-dimensional, are displayed together in Figure \ref{CloudDensity}.

\begin{figure}
\plotone{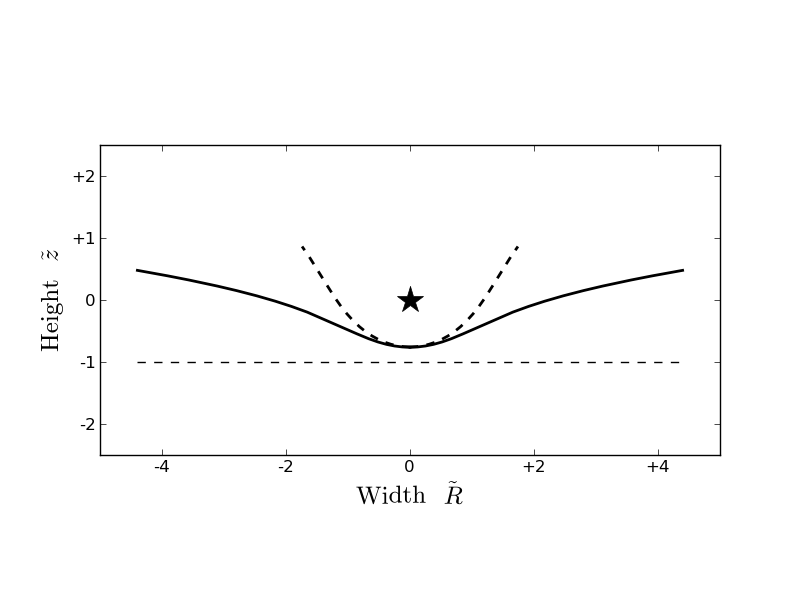}
\caption{Ionization fronts for the slab and pseudo-cylindrical cloud models shown in Figure \ref{CloudDensity}. The horizontal dashed line corresponds to the cloud midplane for both models.}
\label{CylindricalIfront}
\end{figure}

Figure \ref{CylindricalIfront} compares the ionization front shapes for the two cloud models. Because of its relatively high $n_0$-value, the ionization front in this slab model is more flared than in the fiducial one (recall Fig. \ref{IfrontVaryEverything}). The degree of flaring in the pseudo-cylindrical case is much less, a consequence of the gentler falloff in the cloud density profile. 

\begin{figure}
\plotone{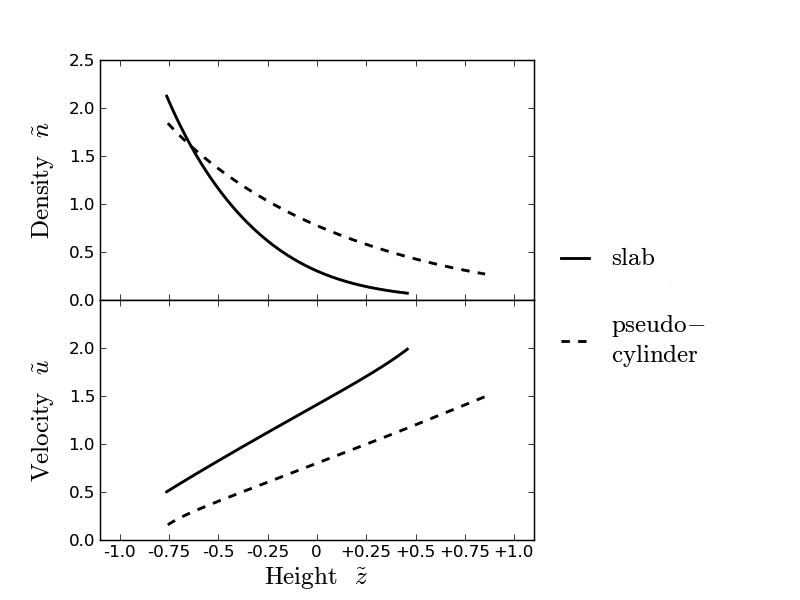}
\caption{Density and velocity profiles for the slab and pseudo-cylindrical cloud models shown in Figure \ref{CloudDensity}}
\label{DensityAndVelocityVaryGeometry}
\end{figure}

Finally, Figure \ref{DensityAndVelocityVaryGeometry} compares the density and velocity profiles within the flows themselves. The flow density within the pseudo-cylinder starts out lower and falls off more slowly. Since the thermal pressure gradient is reduced, so is the acceleration of the ionized gas. As seen in the lower panel of the figure, the velocity begins at a more subsonic value and thereafter climbs less steeply.

\section{Comparison to Observations}
\label{ComparisonToObservations}

\subsection{Emission measure maps}
\label{EmissionMeasureMaps}

\begin{figure}
\epsscale{0.5}
\plotone{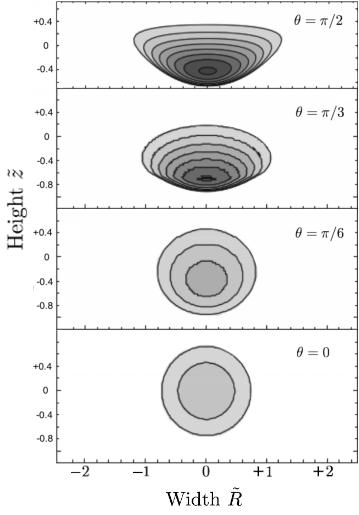}
\caption{Emission measure contours for our fiducial UCHII model viewed at various angles $\theta$ with respect to the flow's central $z$-axis. The peak emission measure is $4.7 \times 10^7 \, \, {\rm cm}^{-6} \, \, {\rm pc}$, and the contours in the top panel represent 0.95, 0.85, ... 0.15 times that value. A grey scale has been applied consistently to all the maps to indicate which emission measures correspond to the plotted contours.}
\label{EMcontours}
\end{figure}

One way to compare our model with observations is to generate synthetic contour maps of the radio continuum emission measure. The latter is the integral of $n^2$ with respect to distance along each line of sight that penetrates the outflow. The resulting maps may also be compared to those generated by other theoretical models, such as the ones of \citeauthor{Redman1998} (\citeyear{Redman1998}; Fig. 3) and \citeauthor{Arthur2006} (\citeyear{Arthur2006}; Fig. 7), 

Figure \ref{EMcontours} displays a series of emission measure maps at different viewing angles, all using our fiducial model. Here, the inclination angle $\theta$ is that between the flow's central ($z$-) axis and the line of sight. The star, as always, lies at the origin, and the spatial coordinates are non-dimensional.

For a flow oriented in the plane of the sky ($\theta = \pi/2$), the brightest emission occurs near the base of the flow, where the ionized gas is relatively dense. The precise location of the peak emission point depends on two competing factors. The width of the ionized flow monotonically increases with $z$, while the $n^2$ decreases. In practice, the peak emission is located about midway between the flow base and the star. The latter two points are separated by a physical distance of about $Z_{49}/2$ (recall eq. (\ref{FiducialLength})). 

The peak emission measure for our fiducial model is $4.7 \times 10^7$ cm$^{-6}$ pc. When ${\cal N}_{\ast}$ is lowered by a factor of 10, as in Table 1, the peak emission measure drops to $1.6 \times 10^7$ cm$^{-6}$ pc. Suppose, on the other hand, we fix $N_\ast$ at its fiducial value while increasing $n_0$ and concurrently decreasing $\zeta$ in the manner described below equation (\ref{MdotScalingSummary}). Then we find that the peak emission measure reaches $1.9 \times 10^8$ cm$^{-6}$ pc for an $n_0$ of $3 \beta /4 = 18.75$. Of the 15 cometaries in \citet{Wood1989} with estimated peak emission measures, the figure varies from $2 \times 10^7$ to $3 \times 10^9$ cm$^{-6}$ pc. Of the 12 cometaries in \citet{Kurtz1994} with estimated peak emission measures, they vary from $1 \times 10^6$ to $6 \times 10^8$ cm$^{-6}$ pc. Thus our model, including reasonable parameter variations, yields peak emission measures that fall within the middle range of observed values.

As the $z$-axis tips toward the line of sight, the emission becomes fainter. The reason is that the total emission measure along any line of sight is increasingly weighted by more rarefied gas at higher values of $z$. Thus, as seen in Figure \ref{EMcontours}, the contours representing higher values of the emission measure successively disappear as $\theta$ decreases. 

For $\theta$ close to zero, the shape of the remaining contours is nearly spherical. However, our synthetic maps fail to reproduce the “shell” or “core-halo” morphologies identified by \citet{Wood1989}. The latter have pronounced limb brightening. As previously noted by \citet{Zhu2005}, champagne-flow models have difficulty explaining this feature.

\begin{figure}
\epsscale{1.0}
\plotone{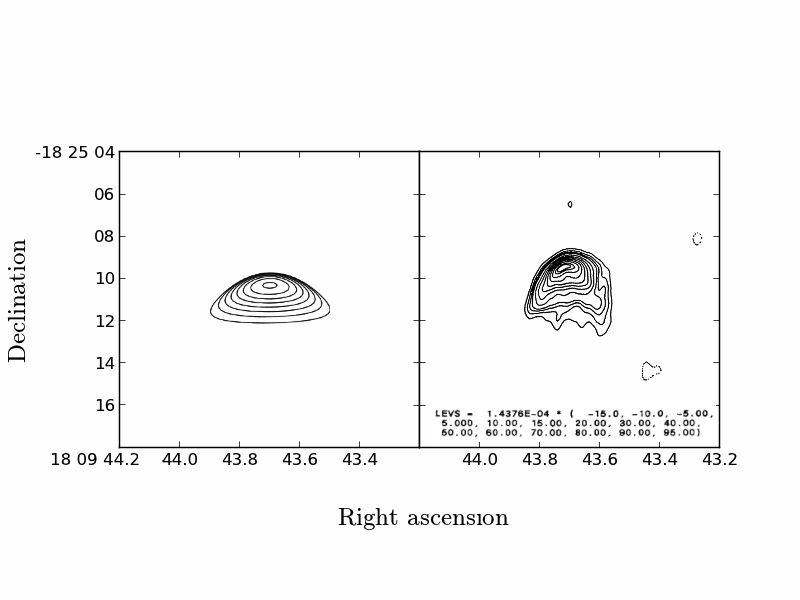}
\caption{ Contours of radio continuum emission.  The left panel uses our fiducial theoretical model, with inclination angle $\theta = \pi/2$ and contours as described in the caption of Figure \ref{EMcontours}. The right panel is an image of G12.21-0.10 from the survey by \citet{Wood1989}, reproduced by the permission of the AAS. The peak emission measure in our model is $4.7 \times 10^7$ pc cm$^{-6}$, while the inferred peak emission measure for G12.21-0.10 is $5.2 \times 10^7$ pc cm$^{-6}$. The reader should refer to \citet{Wood1989} for the flux levels corresponding to each contour.}
\label{RadioComparison}
\end{figure}

Returning to generic cometary regions, our model fits their morphology quite well. Figure~\ref{RadioComparison} compares our fiducial flow (with $\theta = \pi/2$) to an emission map taken from the survey of \citet{Wood1989}. This specific UCHII region has an estimated distance of 16.1 kpc. Accordingly, our synthetic map is displayed in angular coordinates.  We could, in principle, make more precise fits to both this region and many others, by tuning our three free parameters and the viewing angle. Here, we do not attempt such a detailed, comprehensive matching.

\subsection{Velocity structure}

As explained at the outset of Section \ref{DensityAndPotential}, we calculate a laterally-averaged ionized gas velocity at each $z$-location within our model HII region. For this reason, we cannot attempt to match detailed observations of UCHII velocities along particular lines of sight. Nevertheless, we do reproduce several broad characteristics of some UCHII regions. Our model yields a transonic flow that proceeds along the cometary axis, with steep acceleration from head to tail. A similar pattern is seen in the objects observed by \citet{Garay1994}. For our fiducial cometary UCHII region, the velocity gradient we compute is 8 km s$^{-1}$ pc$^{-1}$, about a factor of 2 lower than the gradient measured in G32.80 $\pm$ 0.19. However, as is evident in our Figure \ref{VelocityVaryN0AndZeta}, raising the cloud density and decreasing the displacement of the star from the midplane can give a velocity gradient substantially larger than our fiducial one.

For the bipolar case, we find steady acceleration in both directions along the central axis. \citet{Garay1999} observed this feature in several bipolar compact HII regions. The velocity gradient we compute for our fiducial bipolar model is 50 km s$^{-1}$ pc$^{-1}$, about one third the gradient measured in the compact bipolar HII region K3-50A \citep{Depree1994}.

Like all champagne-flow models, ours predicts that the ionized and molecular gas should be closest in velocity near the head of the cometary region, and that the two velocities should grow farther apart down the tail of the ionized region. This is not the case for G13.87 $\pm 0.28$, for which the ionized velocity closely approaches the molecular cloud velocity in the tail, a characteristic of bowshock models \citep{Garay1994}. However, in G32.80 + .19, the central line emission of the ionized gas in the tail shows accelerates to a speed at least 6 km s$^{-1}$ greater than that of the molecular cloud \citep{Garay1994}. This systematic increase is more consistent with a champagne flow.

For completeness, we note that there exist a number of cometary UCHII regions for which neither a champagne flow nor bowshock accurately accounts for the detailed velocity structure. One well-studied example is G29.96-02 \citep{Martin-Hernandez2003}. Objects such as these have motivated numerical simulations in which a champagne flow is combined with a stellar wind and  motion of the star itself through a molecular cloud \citep{Arthur2006}. 

In summary, a pure champagne flow model such as ours is inadequate to explain all the observed features of cometary UCHII regions. We are encouraged, however, that the main characteristics are well produced. In addition, the simplicity of our quasi-one dimensional model in comparison with multidimensional numerical simulations, allows a much broader exploration of parameter space that will aid in a general understanding of the phenomenon.

\section{Summary and Discussion}
\label{Conclusion}

We have explored a quasi-one dimensional, steady-state wind model for UCHII regions. This picture offers a natural explanation for the cometary morphology that is frequently observed. We are also able to match, at least broadly, such basic observational quantities as the diameter of the region, its peak emission measure, and the ionized gas velocity. Additionally, we have employed the model to see how the density and velocity structure of the region respond to changes in the stellar luminosity, molecular cloud density, and displacement of the exciting star from the peak molecular density. Checking these trends against observations is a task for the future.

In our view, the very young, massive star creating the ionization is still buried deep within within a large molecular cloud. Ionizing photons from the star steadily erode the cloud, expelling gas at a rate of order $10^{-4}$ $M_\odot$~yr$^{-1}$. Consequently, over the inferred UCHII lifetimes of order $10^5$~yr, the star drives out only a relatively small amount of cloud gas, representing the dense clump in which it was born. As time goes on, we expect the flow to grow in size, with the mass expulsion rate concurrently rising.

The ionized gas just inside the ionization front is underpressured with respect to the surrounding neutral material. Indeed, it is just this pressure difference that draws in more neutral material and sustains the wind. \citet{Roshi2005} observe just such a difference in the UCHII region G35.20-1.74. In our model, this pressure difference is \emph{not} due to the generation of a shock by the expanding HII region as in the classical model for D-type fronts (Spitzer \citeyear{Spitzer1978}, Chapter 12). Such a shock probably did form earlier in the evolution of the UCHII region, but has died away by the time a steady flow is established. 

Our semi-analytic model necessarily entails simplifying assumptions. One is our neglect of the stellar wind, which could partially excavate the ionized region. In over half the objects studied in \citet{Zhu2008}, the ionized gas appears to be skirting around a central cavity, possibly indicating the influence of the wind. Note also that a wind would produce an additional shock near the front, elevating the pressure difference.

We have also assumed a steady-state flow in which we neglect the relatively slow advance of the ionization front into the molecular cloud. Recall that equation (\ref{CloudVelocity}) gives $v_\mathrm{cl}^\prime$, the incoming velocity of cloud gas in the frame of the ionization front. Using a typical injection speed of $v_\mathrm{inj}^\prime = (1/2) \, a_I$, along with $n = n_{49}$ and $n_{\rm cl} = (\beta/4) n_{49}$, we find that $v_\mathrm{inj}^\prime = a_I/2 \beta = 0.2$ km s$^{-1}$ , which is indeed small compared to the effective cloud sound speed of 2 km s$^{-1}$. However, over the $10^5$~yr lifetime of the UCHII region, the front advances by 0.02 pc, an appreciable fraction of the region's diameter. A more complete calculation would account for the motion of this front, through retention of the flux leakage term in equation (\ref{RayTracingRewritten}).

In the future, both our type of semi-analytic model and
multidimensional numerical simulations can play valuable roles.
Extending the present calculation, one could follow the spread of
ionization to see how the HII region grows and disperses the bulk of
the molecular cloud. Simulations can eventually provide a more
realistic picture of the parent cloud. In particular, we see a need to
explore the role of turbulence, which we have treated simplistically
as providing an enhanced, isotropic pressure. As \citet{Arthur2011}
have shown, the advance of an HII region into a turbulent cloud is
quite different from the classical account. Within our model, the wind
itself may not be a laminar flow, as we have assumed. Inclusion of
these effects may be necessary to explain the more extreme members of
the irregular UCHII class identified by \citet{Wood1989}.

\section*{Acknowledgements}

Throughout the work, NR was supported by the Department of Energy Office of Science Graduate Fellowship Program (DOE SCGF), made possible in part by the American Recovery and Reinvestment Act of 2009, administered by ORISE-ORAU under contract no. DE-AC05-06OR23100. SS was partially supported by NSF Grant 0908573.

\appendix

\section{Relation between $\epsilon$ and ${\cal N}_\ast$}
\label{GammaNRelation}
Consider a massive star of bolometric luminosity $L_\ast$, radius $R_\ast$, and temperature $T_\ast$. Assuming the star emits like a blackbody, the mean energy of its ionizing photons is 
\begin{eqnarray}
\label{ETrelation}
\epsilon  &=& \frac{\int _{\nu_\mathrm{crit}}^\infty \! B_\nu \, d\nu}{\int_{\nu_\mathrm{crit}}^\infty \! B_\nu/(h \nu) \,  d\nu} \nonumber \\
&=& k\, T_\ast \frac{\int _{x_\mathrm{crit}}^\infty \! x^3/(e^x - 1) \, dx}{\int _{x_\mathrm{crit}}^\infty \! x^2/(e^x - 1) \, dx} \, \, ,
\end{eqnarray}
where $x \equiv h \nu / (k T_\ast)$. Here, $\nu _{\rm crit}$ denotes the Lyman continuum frequency, and $x_{\rm crit} \equiv h \nu_{\rm crit} /(k T_\ast)$. We also may write 
\begin{equation}
{{\cal N}_\ast}  = 4 \pi^2 R_\ast ^2 \int _{\nu_\mathrm{crit}}^\infty \! \frac{B_\nu}{h \nu} \, d\nu \, \, ,
\end{equation}
and so
\begin{equation}
{{\cal N}_\ast} \propto R_\ast ^2 \, T_\ast^3 \, \int _{x_\mathrm{crit}}^\infty \! \frac{x^2}{e^x - 1} \, dx \, \, .
\end{equation}
We next use the standard scaling relations $R_\ast \propto M_\ast^{0.75}$, $L_\ast \propto M_\ast^{3.5}$ for massive main sequence stars \citep[e.g.][]{Hansen2004} as well as the blackbody relation, $L_\ast \propto R_\ast^2 \, T_\ast ^4$. We find
\begin{equation}
\label{NTrelation}
\frac{{{\cal N}_\ast}}{{\cal N}_{49}} = \left(\frac{T_\ast}{3.97 \times 10^4 \, \, {\rm K}}\right)^6 \int _{x_\mathrm{crit}}^\infty \! \frac{x^2}{e^x - 1} \, dx \, \, .
\end{equation}
Here, we have implicitly assumed that the bulk of the star's luminosity is in ionizing photons. We have used results of \citet{Vacca1996} to set the proportionality constant in the last relation. For any $T_\ast$, equation (\ref{NTrelation}) then gives us ${\cal N}_\ast$, while equation (\ref{ETrelation}) gives $\epsilon$. We thus obtain the functional dependence of $\epsilon$ on ${\cal N}_\ast$. Finally, equation (\ref{GammaDefinition}) in the text gives $\gamma$ as a function of $\tilde{{\cal N}_\ast}$, as plotted in Figure \ref{GammaVaryN}.

\begin{figure}
\epsscale{1.0}
\plotone{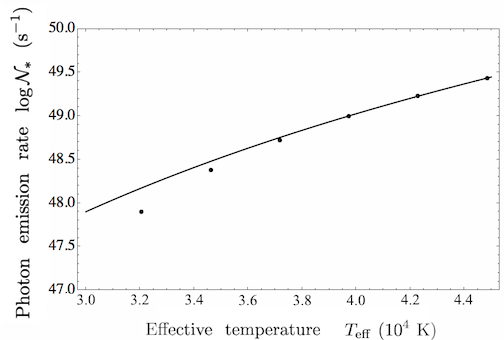}
\caption{The ionizing photon emission rate from an O star as a function of its effective temperature. Shown are discrete values are from \citet{Vacca1996} along with our approximate analytic result (solid curve).}
\label{VaccaPoints}
\end{figure}

As a check on the accuracy of our prescription, Figure \ref{VaccaPoints} shows $N_\ast$ as a function of $T_\ast$, both from our analysis and from the more detailed calculations of \citet{Vacca1996}. Our approximation is most accurate at the highest luminosities, but tends to overestimate the ionizing photon emission rate for the lowest luminosities we consider.

\section{Boundary conditions at base of flow}
\label{BaseBoundaryConditions}
We expand the derivative on the lefthand side of equation (\ref{MassConservation}), and factor out $dR/dz$ from the radical. We thus find  
\begin{equation}
2 \, R \frac{dR}{dz} \, n \, u + R^2 \, \frac{dn}{dz} u + R^2\, n \frac{du}{dz} = 2 \, R \, \frac{dR}{dz} \sqrt{1 + \left(\frac{dz}{dR}\right)^2}  n \, v_\mathrm{inj} \, \, .
\label{BoundaryMassConservation}
\end{equation}
At the base, both $R$ and $dz/dR$ tend to zero. Thus, the second and third terms on the lefthand side of equation (\ref{BoundaryMassConservation}) vanish, as does the square root on the righthand side. We therefore have 
\begin{equation}
u = {v_\mathrm{inj}} \, \, ,
\label{BaseContinuity}
\end{equation}
exactly at the base. Recall that $v_\mathrm{inj}$ is a function of $n$. We choose $n$ at the base such that we obtain a transonic solution via the method of shooting and splitting.

We also need to know the values of $dn/dz$ and $du/dz$ at the base. We begin by rewriting the decoupled, nondimensional equations of motion as
\begin{equation}
(1 - u^2)\frac{du}{dz} - u \frac{d\Phi_{\rm cl}}{dz} + u f_\mathrm{rad} = \frac{2\sqrt{1 + \left(dz/dR\right)^2}(1 + u^2)v_\mathrm{inj} - 2(1 + v_\mathrm{inj}^2 )u}{R \, \, dz/dR}
\end{equation}
\begin{equation}
(1 - u^2)\frac{dn}{dz} + n \frac{d\Phi_{\rm cl}}{dz} - n f_\mathrm{rad} = \frac{2(u^2 + v_\mathrm{inj}^2)n -4 n \sqrt{1 + \left(dz/dR\right)^2} u \, v_\mathrm{inj}}{R \, \, dz/dR} \, \, ,
\end{equation}
where $f_{\rm rad}$ is given by equation (\ref{FradDefinition}). In the limits $R \rightarrow 0$ and $dz/dR \rightarrow 0$, the righthand sides of both equations can be evaluated using L'H$\hat{\rm o}$pital's rule. In order to compute quantities such as $d^2 z/dR^2$, we use the fact that, to lowest order in $R$, the cavity wall is parabolic, i.e. $z \approx A \, R^2 - H_b$ where $A$ is a constant. At each timestep, we perform a numerical fit to determine $A$. The final results for the derivatives at the base are
\begin{equation}
\frac{dn}{dz} = \left[\frac{1}{1 - u^2} \right]\left[ -n \, \frac{d\Phi_{\rm cl}}{dz} +n \, f_\mathrm{rad} - 4 \,A \, n \, u^2\right]
\end{equation}
and
\begin{equation}
\frac{du}{dz} = \left[\frac{1}{1 - u^2} \right]\left[ \frac{1}{2} u \, \frac{d\Phi_{\rm cl}}{dz} - \frac{1}{2} \,u \, f_\mathrm{rad} + \,A \, u (1 + u^2) + \frac{1}{2} \,\frac{dv_\mathrm{inj}}{dz} \, (1-u^2) \right] \, \, .
\label{DudzBase}
\end{equation}
We obtain the term $dv_{\rm inj} / dz$ in equation (\ref{DudzBase}) by differentiating equation (\ref{InjectionVelocityNondimensional}): 
\begin{equation}
\frac{dv_\mathrm{inj}}{dz} = \frac{2(\beta \,n^2 - 8 \, n \, n_\mathrm{cl} + 16 n_\mathrm{cl}^2)(n \, d n_\mathrm{cl}/dz - n_\mathrm{cl} \, dn/dz )}{\beta \, n^2(n - 4 n_\mathrm{cl})^2 v_\mathrm{inj}} \, \, .
\end{equation}

\begin{table*}
{
\centering
\caption{Mass, momentum and energy transport rates} 
\begin{tabular}{cccccc}
\hline
${\cal N}_\ast$ & $\zeta$ & $n_0$ & $\dot{M}_{\rm out}$         & $\dot{p}_{\rm out}$ & $\dot{E}_{\rm out} $ \\
                &         &       & $(M_\odot \, \rm{yr}^{-1})$ & $(\rm{dyne})$       & $(\rm{erg} \, \rm{s}^{-1})$ \\        
\hline
0.1 & 1.0& $\beta/4$ & $6.2 \times 10^{-5}$ & $5.3 \times 10^{27}$ & $3.6 \times 10^{33}$ \\
0.3 & 1.0& $\beta/4$ & $1.6 \times 10^{-4}$ & $1.6 \times 10^{28}$ & $1.3 \times 10^{34}$ \\
1.0 & 1.0 & $\beta/4$ & $3.7 \times 10^{-4}$ & $3.9 \times 10^{28}$ & $3.2 \times 10^{34}$ \\
\hline
1.0 & 1.5 & $\beta/4$ & $6.0 \times 10^{-4}$ & $6.8 \times 10^{28}$ & $6.0 \times 10^{34}$ \\
1.0 & 2.0 & $\beta/4$ & $8.5 \times 10^{-4}$ & $9.9 \times 10^{28}$ & $9.2 \times 10^{34}$ \\
\hline
1.0 & 1.0 & $\beta/2$ & $2.4 \times 10^{-4}$ & $2.7 \times 10^{28}$ & $2.3 \times 10^{34}$ \\
1.0 & 1.0 & $3\beta/4$ & $4.7 \times 10^{-4}$ & $5.6 \times 10^{28}$ & $5.2 \times 10^{34}$ \\
1.0 & 1.0 & $\beta$ & $5.3 \times 10^{-4}$ & $6.5 \times 10^{28}$ & $6.3 \times 10^{34}$ \\
\hline
1.0 & $\sqrt{3/2}$ & $\beta/6$ & $4.3 \times 10^{-4}$  & $4.4 \times 10^{28}$ & $3.6 \times 10^{34}$ \\
1.0 & $\sqrt{1/2}$ & $\beta/2$ & $2.9 \times 10^{-4}$  & $3.0 \times 10^{28}$ &  $2.5 \times 10^{34}$ \\
1.0 & $\sqrt{1/3}$ & $3\beta/4$ & $2.5 \times 10^{-4}$ & $2.6 \times 10^{28}$ & $2.2 \times  10^{34}$ \\
\end{tabular}
}
\label{MassLossTable}
\end{table*}

\clearpage

\clearpage

\clearpage

\clearpage

\clearpage

\clearpage

\clearpage

\clearpage

\clearpage

\clearpage

\clearpage

\clearpage

\clearpage

\clearpage

\clearpage

\clearpage

\clearpage

\clearpage

\clearpage

\clearpage

\clearpage

\end{document}